\documentclass[11pt]{article}
\pdfoutput=1
\usepackage[utf8]{inputenc}
\usepackage{multirow}
\usepackage{svg}
\usepackage{array}
\usepackage{amsmath, amsfonts, amssymb}
\usepackage{comment}
\usepackage{graphicx}
\usepackage{psfrag}
\usepackage{amsthm}
\usepackage{xcolor}
\usepackage{enumerate}
\usepackage{arydshln}
\usepackage{soul}
 \usepackage{slashed}
\usepackage{mathrsfs}
 \usepackage{a4wide}
  \usepackage{tikz}
  \usepackage{tikz-cd}
  \usetikzlibrary{shapes.geometric}
  \usepackage{tcolorbox}
\usepackage{wrapfig}
  \usepackage{color}
  \definecolor{dark-gray}{gray}{0.20}
  \definecolor{gray}{gray}{0.30}
  \definecolor{light-gray}{gray}{0.80}
  \definecolor{dark-red}{rgb}{0.7,0,0}
  \definecolor{dark-green}{rgb}{0.1,0.4,0}
  \definecolor{dark-blue}{rgb}{0.3,0.3,0.7}
  \definecolor{light-blue}{rgb}{0.8,0.8,1}
      \definecolor{swamp}{RGB}{240, 199, 197}

  \usepackage{pifont}

\usepackage{newunicodechar} 

\usepackage{setspace}

\newcommand{\be}{\begin{equation}}
\newcommand{\ee}{\end{equation}}

\def\be{\begin{equation}}
\def\ee{\end{equation}}
\def\bea{\begin{eqnarray}}
\def\eea{\end{eqnarray}}

\definecolor{antiquefuchsia}{rgb}{0.57, 0.36, 0.51}

\newcommand{\tl}{\mathbf{t}}

\def\simleq{\; \raise0.3ex\hbox{$<$\kern-0.75em
      \raise-1.1ex\hbox{$\sim$}}\; }
   \def\simgeq{\; \raise0.3ex\hbox{$>$\kern-0.75em
      \raise-1.1ex\hbox{$\sim$}}\; }

\numberwithin{equation}{section}

\usepackage{tocloft}

\usepackage{subfig}

\usepackage{jheppub}
\usepackage{hyperref}
\usepackage{cleveref}

\usepackage{color}

\definecolor{dark-blue2}{RGB}{0, 114,178}

\hypersetup{
	colorlinks=true,
	linkcolor=dark-blue2,
	citecolor=dark-blue2,
	urlcolor=dark-blue2,
	linktoc=section
}

\theoremstyle{remark}

\crefname{appendix}{Appendix}{Appendices}

\def\({\left(}
\def\){\right)}
\def\[{\left[}
\def\]{\right]}

\title{\centering  Classical Black Hole Probes of UV Scales}

\author{Jos\'e Calder\'on-Infante$^1$,}
\author{Matilda Delgado$^{2,3}$,}
\author{Yixuan Li$^{4,5}$,}
\author{Dieter L\"ust$^{2,6}$,}
\author{Angel Uranga$^7$}
\affiliation{$^{1}$CERN, Theoretical Physics Department, 1211 Meyrin, Switzerland}
\affiliation{$^2$Max-Planck-Institut f\"ur Physik (Werner-Heisenberg-Institut)\\
Boltzmannstr. 8, 85748 Garching, Germany}
\affiliation{$^3$Jefferson Physical Laboratory, Harvard University\\ 17 Oxford St, Cambridge, MA 02138, United States of America}
\affiliation{$^4$Dipartimento di Fisica e Astronomia “Galileo Galilei”, Università di Padova, Via Marzolo 8, 35131 Padova, Italy}
\affiliation{$^5$INFN, Sezione di Padova, Via Marzolo 8, 35131 Padova, Italy}
\affiliation{$^6$Arnold Sommerfeld Center for Theoretical Physics\\ Ludwig-Maximilians-Universit\"at M\"unchen, 80333 M\"unchen, Germany}
\affiliation{$^7$Instituto de F\'{i}sica Te\'{o}rica IFT-UAM/CSIC\\
C/ Nicol\'{a}s Cabrera 13-15, Campus de Cantoblanco, 28049 Madrid, Spain}

\emailAdd{jose.calderon-infante@cern.ch}
\emailAdd{matilda@mpp.mpg.de}
\emailAdd{yixuan.li@pd.infn.it}
\emailAdd{luest@mpp.mpg.de}
\emailAdd{angel.uranga@csic.es}

\abstract{In the context of the Swampland program, black hole attractors have been employed to probe infinite distances in moduli space, where the EFT cutoff goes to zero in Planck units and UV effects become significant. In this paper, we take the perspective of the two-derivative action of string theoretic effective field theories and explore various families of extremal black hole solutions that probe infinite distance limits at their horizons. While these solutions do not include higher-order corrections in the EFT expansion, we find that, in many cases, the smallest BPS black holes in these families remarkably reproduce either the species scale or some other Kaluza-Klein scale. In highly supersymmetric cases, this match with UV scales even persists in the interior of moduli space. We explore the use of the species length bound on classical black hole sizes as a criterion to rule out inconsistent EFTs, using a 9d bottom-up model arising from an  {\em ad hoc} truncation of maximal supergravity. These observations suggests that the two-derivative action may encode information about relevant UV scales. We discuss the interplay of these results with emergence and UV/IR mixing in quantum gravity.
}

\begin{document}
\hypersetup{pageanchor=false}
\makeatletter
\let\old@fpheader\@fpheader
\preprint{CERN-TH-2025-033, MPP-2025-10, LMU-ASC 04/25, IFT-UAM/CSIC-25-009}

\makeatother

\maketitle

\section{Introduction}
\label{sec:intro}

One of the main recent outcomes of the Swampland program \cite{Vafa:2005ui,Ooguri:2006in} (see \cite{Palti:2019pca,vanBeest:2021lhn,Grana:2021zvf,Agmon:2022thq} for reviews), is the enormous progress surrounding the classification of infinite distance limits in the moduli space of effective field theories (EFTs) that come from quantum gravity  (see \cite{Klaewer:2016kiy,Grimm:2018ohb,Ooguri:2018wrx,Corvilain:2018lgw,Grimm:2018cpv,Buratti:2018xjt,Marchesano:2019ifh,Lee:2019xtm,Lust:2019zwm,Kehagias:2019akr,Lee:2019wij,Baume:2019sry,Bonnefoy:2019nzv,Gendler:2020dfp,Calderon-Infante:2020dhm, Baume:2020dqd,Perlmutter:2020buo,Cribiori:2021gbf,Castellano:2021yye,Buratti:2021fiv,Etheredge:2022opl,Angius:2022aeq,Angius:2023xtu,Baume:2023msm,Etheredge:2023odp,Calderon-Infante:2023ler,Castellano:2023jjt,Castellano:2023stg,Basile:2023blg,Demulder:2023vlo,Angius:2023uqk,Angius:2024zjv,Etheredge:2024tok,Calderon-Infante:2024oed,Demulder:2024glx}, for related developments). In these limits, the EFT description breaks down due to infinite towers of states, which become exponentially light according to the Distance Conjecture \cite{Ooguri:2006in}. One can identify various physical scales that become light in these limits, and which have been proposed \cite{Lee:2019wij} to always correspond to either Kaluza-Klein (KK) modes of decompactifying extra dimensions or the excitations of an emergent weakly-coupled string. 

The lowering of the UV cutoff of the EFT by a large number of light modes is efficiently captured by the introduction of the species scale (see \cite{Dvali:2007hz,Dvali:2007wp,Dvali:2008ec,Dvali:2009ks,Dvali:2010vm,Dvali:2012uq} for early references and \cite{Castellano:2022bvr,vandeHeisteeg:2022btw,Cribiori:2022nke,Cribiori:2023ffn,Cribiori:2023sch,Calderon-Infante:2023uhz,Basile:2024dqq,Herraez:2024kux,Anastasi:2025puv,ValeixoBento:2025iqu,Calderon-Infante:2025ldq} for recent explorations in the Swampland context). This has been alternatively characterized as the scale controlling certain higher-curvature corrections to the effective action \cite{vandeHeisteeg:2022btw,Cribiori:2022nke,vandeHeisteeg:2023ubh,Calderon-Infante:2023uhz, Castellano:2023aum, vandeHeisteeg:2023dlw, Bedroya:2024ubj, Aoufia:2024awo,Castellano:2024bna} (see \cite{Calderon-Infante:2025ldq} for a recent discussion). The species scale corresponds to the energy scale at which quantum gravitational effects become manifest. 

It becomes both a fundamental and phenomenological challenge to investigate how the infinite distance limits of moduli space might be probed within the constraints of a finite-sized experiment. In this context, charged black holes in string-theoretic EFTs play a pivotal role. Indeed, in such setups the gauge couplings of the U(1) symmetries, under which these black holes are charged, typically depend on the moduli. As a result, the presence of such black holes induces an effective potential for the moduli fields, dynamically attracting them to specific values at the black hole horizon. This is what is known as the black hole \textit{attractor mechanism}, by which the values of the moduli at the horizon are entirely determined by the black hole charges. Supersymmetric versions of such black holes have already been used at various instances to push the moduli to large values and probe infinite distance limits. For instance, it was discussed in \cite{Bonnefoy:2019nzv} that a tower of light states is expected to emerge in the limit of large black hole entropy. Moreover, it was argued in \cite{Delgado:2022dkz} that such black holes encode UV data about the underlying compactification in their thermodynamic properties. Both of these examples showcase how such black holes can be used as IR probes of UV effects. 

The use of black holes as probes of the UV is further motivated by the original definition of the species scale as the size of the smallest possible, the so-called \textit{minimal} black hole, describable in the EFT \cite{Dvali:2007hz,Dvali:2007wp,Dvali:2008ec,Dvali:2009ks,Dvali:2010vm,Dvali:2012uq}. In fact, the sensitivity of minimal black holes to the species scale has been further argued to follow from the assertion \cite{Cribiori:2023ffn,Basile:2023blg,Basile:2024dqq,Herraez:2024kux} that minimal black holes can be viewed as bound states of the KK modes or emergent string excitations that become light in these limits as per the Distance Conjecture \cite{Ooguri:2006in}. This observation is connected to tower-black hole transitions \cite{Herraez:2024kux}.

To this effect, in the context of supersymmetric theories the BPS attractor mechanism for black 
holes in 4d $\mathcal{N}=2$ \cite{Ferrara:1995ih,Ferrara_1996a,Ferrara_1996b,Ferrara_1997,moore,Moore:1998zu,Strominger:1996kf,Denef_1999,Denef:2001xn} and 5d $\mathcal{N}=1$ \cite{Kraus:2005gh,Larsen:2006xm} has been used at various instances e.g. \cite{Cribiori:2022nke,Cribiori:2023swd,Cribiori:2023ffn,Calderon-Infante:2023uhz,Basile:2024dqq,Li:2021gbg,Li:2021utg} to drag moduli to large distances in moduli space. Most such explorations consider 
singular (i.e. zero horizon size) black holes, with horizons stretched due to higher curvature corrections (see e.g. \cite{Dabholkar:2012zz} for a review). Others have computed how curvature corrections affect BPS black holes in the asymptotic limit, tracking how the solution gets uplifted to a higher-dimensional one in detail \cite{Castellano:2025ljk}.  Although these approaches have provided great insights into the link between the various definitions of the species scale (i.e. as the size of the minimal black hole, or the scale controlling certain higher-curvature corrections), they rely on knowing about those corrections, which are typically computable only in the full UV complete theory. 

In this paper, we explore a complementary approach that probes the structure of UV scales of the EFT using only low-energy data. In particular, we explore the UV scales by using the charged black hole solutions in the 2-derivative theory, and picking the smallest ones, namely the \textit{minimal black holes within this classical approximation}. In this work we will simply dub them {\em minimal black holes}, although the theory might a priori have even smaller ones arising from quantum stretched horizons as mentioned above. Indeed, we propose to study these naive two-derivative solutions in complete disregard of the curvature corrections that may affect these minimal black holes. Because of this, one would therefore not expect these classical black hole solutions to be sensitive to UV physics. We nevertheless consider the extrapolation of these solutions in the limit where their horizon become as small as possible while exploring the boundary of moduli space, and find that such classical minimal black holes do follow a relevant UV scale of the theory. In other words, one cannot make these classical black holes as small as one wants; their size is bounded from below by UV cutoff scales in the theory. This generalizes the observation in \cite{Cribiori:2023ffn} that the sizes of specific families of classical minimal black holes in 4d $\mathcal{N}=2$ and 4d $\mathcal{N}=4$ agree with the species scale in the boundary of moduli space. This is surprising, as one would expect these classical, two-derivative black hole solution to be blind to such a UV scale. This result prompts us to investigate whether it is a general phenomenon that such classical black holes ``know'' about UV scales in this way. To this effect, we carry out an exploration of families of minimal black holes whose attractor values for the moduli explore infinite distance limits. We aim to explore how their sizes compare to physically meaningful scales, such as the KK scale or the species scale.

More specifically, we test these ideas using 4d and 5d BPS black hole attractors solutions in string theoretic EFTs. For each different infinite distance limit in vector multiplet moduli space, we ask whether there is a class of \emph{classical} black holes that follows the species scale. We find that in most cases, the answer is yes: there is always a family of classical minimal black holes that follows the species scale asymptotically. Moreover, in some cases, it can be shown under reasonable assumptions that there are no BPS black holes parametrically smaller than the species scale in these limits (so all black holes of this kind are bigger than this general bound). In addition, we present partial results that this behavior persists in the interior of moduli space. These findings are suggestive that the 2-derivative theory knows about the species scale, a possibility which tantalizingly resonates with the concept of emergence (see e.g. \cite{Heidenreich:2017sim,Grimm:2018ohb,Heidenreich:2018kpg,Palti:2019pca,Marchesano:2022axe,Castellano:2022bvr,Castellano:2023qhp,Blumenhagen:2023yws,Blumenhagen:2023tev,Blumenhagen:2023xmk,Hattab:2023moj,Blumenhagen:2024ydy} for proposals of emergence in the swampland context). This establishes a remarkable connection between IR physics, manifested through solutions to the supergravity two-derivative action, and UV scales. Our work thus represents a first step toward exploring the possibility of probing these UV scales using classical black holes, offering a promising new realization of UV/IR mixing.

That being said, this connection does not hold generally and there are subtleties. For instance, we also find a case where one can show that no BPS black hole follows the species scale, and where the minimal black holes follow instead a KK scale. Although it does not follow the species scale, it still suggests that the two-derivative action knows about the UV. We provide heuristic explanations to distinguish between the two situations, based on considerations surrounding the microstates of the black hole and the states in the tower becoming light in that limit.

However, we prove that in all cases the \textbf{classical black holes are never smaller than the species length in the different asymptotic limits}. We propose that this is a general feature of consistent theories, potentially providing a swampland criterion to rule out inconsistent EFTs. We explore this proposal in a 9d bottom-up toy model obtained from an {\em ad hoc} truncation of 9d maximal supergravity, where we show that the UV-complete theory has no finite-size classical black holes violating our proposal, while the bottom-up toy model does. \\

We now summarize our results, which are also organized in Table \ref{fig:enter-label}. First, we consider 5d ${\cal N}=1$ theories from Calabi-Yau threefold compactifications of M-theory, and explore different asymptotic limits, corresponding to either decompactification or emergent string limits. We find that in all limits the UV scale reproduced by minimal classical black holes precisely agrees with the species scale. We moreover show in an illustrative example that this agreement persists at the semiquantitative level even in the interior of moduli space.

Then, we consider 4d $\mathcal{N}=2$ compactifications of type II string theory, where there are three different types of infinite-distance limits \cite{Grimm:2018ohb,Corvilain:2018lgw,Lee:2019wij}. Adopting the notation in \cite{Grimm:2018ohb,Corvilain:2018lgw}, we dub them Type IV, III and II limits. We construct several infinite families of classical black holes in the STU model, which can explore those asymptotic regimes, and we determine the UV scale probed by their minimal representatives in those limits. We show that in Type IV and Type II limits, which correspond to a decompactification and an emergent string limit, respectively, the minimal black holes follow the species scale, while for Type III the minimal black holes follow the KK scale. We provide heuristic arguments explaining these results, based on the interplay of the diverging charges with the states in the asymptotic towers. Going beyond the STU model, we can prove for a general Calabi-Yau compactification that there are no BPS black hole solutions that become parametrically smaller than the species scale in limits where one modulus becomes large. Finally, we study a family of solutions with enhanced supersymmetry (4d $\mathcal{N}=4$) and show that they follow the species scale at the semiquantitative level in the interior of moduli space.

We complete our discussion by considering attractor solutions in a 9d bottom-up model obtained from 9d maximal supergravity by truncating away its axion. The model is simple enough to be exhaustive about the various infinite distance limits there are to probe. We show that the infinite distance limits and UV scales probed by these classical black holes can be organized in a convex hull description, analogous to that in the Species Scale Distance Conjecture \cite{Calderon-Infante:2023ler}. Interestingly, in this case there exist infinite distance limits for which the classical minimal black holes are parametrically smaller than the putative species length, in stark contrast with our proposal. We solve the puzzle by showing that these black holes do not survive in the full 9d supergravity due to the presence of the axion. This suggests that the presence of axions in the EFT, which is crucial for it to actually admit a UV completion in terms of type II string theory on $\mathbb{S}^1$, is also crucial to avoid the presence of classical finite-sized black holes violating the species length bound. This provides further evidence for our proposal: the only black holes that become smaller than the species length in asymptotic limits are  solutions to theories that are in the swampland. This suggests a tantalizing application of our proposal, to use classical minimal black hole sizes as a swampland criterion to rule out inconsistent EFTs.

The above results suggest interesting UV/IR links arising in the 2-derivative EFT and its black hole solutions. We provide some suggestive arguments relating our findings to emergence of the gauge kinetic terms, linking the structure of tree level gauge kinetic functions to the asymptotic towers in infinite distance limits. We also provide microstate counting arguments relating the composition of the charged black hole solutions to the structure of such towers and the species scale.

\begin{figure}
\renewcommand{\arraystretch}{1.3}
\label{table}
    \centering
    \begin{tabular}{||p{3cm}|p{5.5cm}|c||}
    \hline
   \textbf{Theory}  &  \textbf{Limit Type} &\textbf{ Minimal BHs follow the... }\\
   \hline\hline
    \hyperref[sec:5d]{M-theory on CY3} & Decompactification & Species scale\\
   $\;$ & Emergent String & Species scale\\ \hline
   \hyperref[sec:4d]{Type II on CY3} & Emergent String (Type II) & Species scale\\
  $\;$ & Decompactification (Type III) & KK scale\\
 $\;$  & Decompactification (Type IV) & Species scale\\
   \hline

\end{tabular}

    \caption{The summary of our non-exhaustive search for classical BPS black hole families that probe infinite distance limits in moduli space. The classification of the infinite distance limits in CY compactifications of type II string theories is that of \cite{Corvilain:2018lgw}.}
    \label{fig:enter-label}
\end{figure}

In summary, our work opens up the new exciting direction of using classical black holes as probes of UV physics. There are clearly many questions, to which we hope to come back in the future.

\medskip

The paper is organized as follows. In section \ref{sec:5d} we focus on 5d ${\cal N}=1$ theories from M-theory on Calabi-Yau threefolds. We describe the structure of the 5d theory and the set of electrically charged classical black holes in section \ref{sec:5d-background}. In section \ref{sec:5d-asymptotic} we describe the different asymptotic limits in moduli space (section \ref{sec:5d-limits}), the minimal black holes exploring them (section \ref{sec:5d-BH}), and show that the UV scale they probe precisely agrees with the species scale, both in the decompactification and the emergent string limits. In section \ref{sec:5d-interior} we explore the extension of these results to the interior of moduli space. 

In section \ref{sec:4d} we carry out the analysis for 4d ${\cal N}=2$ theories. We describe the 4d theory in section \ref{sec:4d-background}, including its asymptotic limits (section \ref{sec:4d-limits}) and three families of charged black holes (section \ref{sec:4d-limits}), solutions to the STU model. In section \ref{sec:4d-limits-BH} we use the minimal black holes in those families to explore the UV scale they probe along Type IV, III and II limits (sections \ref{sec:4d-limits-BH-iv}, \ref{sec:4d-limits-BH-iii} and \ref{sec:4d-limits-BH-ii}, respectively). In section \ref{sec:lowerbound}, we argue that even beyond the STU model, there cannot be black hole solutions that become parametrically smaller than the species scale in Type IV and II limits (sections \ref{sec:proofIV} and \ref{sec:proofII}, respectively) or the KK scale in Type III limits (section \ref{sec:proofIII}). Finally, we compare minimal black holes sizes and the species scale in the interior of moduli space using a 4d $\mathcal{N}=4$ theory in section \ref{sec:4d-interior}.

In Section \ref{sec:9d} we carry out a complementary analysis and show the potential of our proposal to rule out inconsistent theories. We study a bottom-up 9d effective field theory, which can be regarded as a truncation of maximal 9d supergravity, by removing the axion. The theory has two saxion moduli and is very tractable, allowing for a systematic exploration of the classical black hole spectrum, introduced in section \ref{sec:class-9DBHs}. In section \ref{sec:9d-limits-BH} we use the minimal black holes to explore the UV scale they probe at infinite distance limits, including an emergent string limit (section \ref{sec:9d-limits-BH-string}) and a decompactification to 11d limit (\ref{sec:9d-limits-BH-decomp}). In section \ref{sec:global picture} we provide a global description of the possible infinite distance limits and describe the corresponding distance conjecture convex hull in section \ref{sec:global-distance-conjecture-hull}. In section \ref{sec:global-black-hole-hull} we propose a black hole convex hull diagram to encode the UV scales probed by minimal black holes in general limits. In this setup, the black hole families reproduce the correct species scale in the KK limit, but fail to follow the species scale in the emergent string limit. Even worse, the minimal black holes in the latter families turn out to be parametrically smaller than the species length, in sharp contradiction with our proposal. In section \ref{sec:puzzle-resolution} we solve this puzzle, and show that the axion of the full 9d supergravity are actually active in such a way that the black holes of the truncated theory do not survive its inclusion. This tantalizingly suggests that the fact that classical black holes are smaller than the species scale signals the inconsistency of the theory with the {\em ad hoc} truncation of the axion.

In section \ref{sec:interpretation} we propose some heuristic interpretations of the appearance of UV scales, concretely the species scale, in the physics of black hole solutions of the 2-derivative EFT. Section \ref{sec:tracking} explores this explanation in terms of emergence of the gauge kinetic terms, and their relation to the asymptotic towers of states in infinite distance limits. In section \ref{sec:micro} we relate the size of minimal black holes (equivalently, their entropy) to the counting of microstates using the  number of tower state species available in the theory. Finally, we offer some final remarks in section \ref{sec:conclusions}.

\section{5d $\mathcal{N}=1$ from M-theory on CY3}
\label{sec:5d}

In this first section, we consider classical BPS Black Holes in 5d $\mathcal{N}=1$ compactifications of M-theory. This example will serve as a prototype for the others that follow in Sections \ref{sec:4d} and \ref{sec:9d}. Indeed, the special structure of vector multiplet moduli space in these theories makes them simple enough to be exhaustive about the infinite distance limits that are probed by classical black holes. For our purposes, there are only two qualitatively different types of limits, and the black hole attractor equations can be readily solved. 

We will build families of classical black hole solutions that, through the attractor mechanism, drag the moduli to a specific value at their horizon. We will show that the smallest of such black holes reproduce the species scale in the two different types of infinite-distance limits that exist in the theory. 

Furthermore, in such compactifications, the prepotential is known even in the bulk of moduli space, which allows us to test whether this matching between the smallest black holes and the species scale persists inside the bulk of moduli space. For this, we will consider the proposal in \cite{vandeHeisteeg:2023dlw} (see also \cite{vandeHeisteeg:2022btw,Cribiori:2022nke,vandeHeisteeg:2023ubh,Calderon-Infante:2023uhz, Castellano:2023aum, Bedroya:2024ubj, Aoufia:2024awo,Castellano:2024bna,Calderon-Infante:2025ldq}), where the species scale was defined in the interior of moduli space in relation to the curvature-squared correction to Einstein gravity. We will find that the match is precise in the asymptotic limit, and persists at a semi-quantitative level (i.e. up to ${\cal O}(1)$ factors) in the interior of moduli space.

We start by reviewing 5d $\mathcal{N}=1$ compactifications of M-theory in section \ref{sec:5d-background}. In particular, we describe a general family of BPS black holes which feature an attractor mechanism for the vector multiplet moduli. In section \ref{sec:5d-asymptotic}, we then proceed to using these black holes to probe infinite distance limits in vector multiplet moduli space. We first review the classification of asymptotic limits in section \ref{sec:5d-limits}. All different types of limits can be associated to a fibration structure of the CY3, which allows us to describe the different limits and their associated species scale in a very general manner. Using this general framework, we find the smallest possible BPS black holes families that probe each different type of asymptotic limit in section \ref{sec:5d-BH}. Remarkably, we find that these black holes follow the species scale. Finally, in section \ref{sec:5d-interior}, we examine the size of minimal black holes as one probes the bulk of moduli space. We find that they surprisingly give a reasonably good estimate of the species scale even deep into the bulk. 

\subsection{Structure of the 5d Theory and Black Holes}
\label{sec:5d-background}

In this section we review the structure of the 5d supergravity theory and of the black holes we consider. We follow conventions in \cite{Larsen:2006xm}.

We consider 5d $\mathcal{N}=1$ theories, with M-theory on CY3 in mind. The vector multiplet moduli space of M-theory on a CY 3-fold ${\mathbf X}_6$ with hodge numbers $\{h_{2,1},h_{1,1}\}$ is composed of $h_{1,1} - 1$ vector multiplets (the graviphoton goes into the gravity multiplet). The K\"ahler moduli are the coordinates of the K\"ahler form, $J$, on a basis of the $h_{1,1}$ $(1,1)$-cycles $\Omega^I$:
\begin{equation}
x^I=\int_{\Omega^I} J \quad ; \quad I=1,\ldots, h_{1,1} \,.
\end{equation}
There is a dual basis $\Omega_J$ of four-cycles, and we define the following “dual” period:
\begin{equation}
x_J=\int_{\Omega_J} J\wedge J \,.
\end{equation}
The triple intersection numbers of the CY3 are given by:
\begin{equation}
C_{IJK}=\int_{{\mathbf x}_6} J_I\wedge J_J\wedge J_K \,.
\end{equation}
One of the vector fields is the graviphoton which does not belong to a vector multiplet. Correspondingly, one of the scalars is not in a vector multiplet, but in a hypermultiplet. It corresponds to the overall volume of ${\mathbf X}_6$
 \begin{equation}
\mathcal{V}=\frac{1}{3!}C_{IJK}x^I x^J x^K \,.
\label{volume-5d}
\end{equation}
Hypermultiplets decouple from the attractor mechanism so (\ref{volume-5d}) can be treated as a constraint and we only consider the $h_{1,1} - 1$ unconstrained scalars that are left over. An easy choice is to pick $\mathcal{V} = 1$ , in which case we denote the leftover moduli by $\varphi_i$, with $i = 1, \ldots, h_{1,1} - 1$.

The overall volume (\ref{volume-5d}) plays the role of a prepotential in 4d $\mathcal{N} = 2$ language. We can write the metric on scalar field space as:
\begin{equation}\label{eq:metricon}
g_{ij}=G_{IJ}\partial_i x^I\partial_j x^J\quad ,\quad G_{IJ}=-\frac 12 \partial_I\partial_J\log {\mathcal V}(x)\, ,
\end{equation}
and we have the following relations
\begin{equation}
x_J=\frac 12 C_{IJK} x^J x^K\quad ,\quad G_{IJ}x^J=\frac{1}{2{\mathcal {V}}}x_I \,.
\end{equation}

We can consider BPS black holes with asymptotically flat space in these 5d theories, which have an attractor mechanism for the vector-multiplet moduli. For a black hole with electric charges $Q_ I$ the central charge is given by
\begin{equation}\label{eq:Ze}
Z_e=x^IQ_I \,.
\end{equation}
The attractor equations descibe the extremization of the central charge at the horizon:
\begin{equation}
\partial_i Z_e |_h=0 \,.
\end{equation}
The attractor flow is such that the central charge starts at a maximum in asymptotically flat space and decreases as the black hole is approached. The value of the moduli and the black hole entropy at the extremum are given by
\begin{eqnarray}
x_I\big|_{h}&=&\frac{Q_I}{(\frac{1}{3!}C^{JKL}Q_JQ_KQ_L)^{1/3}}\nonumber \\
S &\sim & \big(Z_e\big|_h\big)^{3/2}=\sqrt{\frac{1}{3!}C^{JKL}Q_JQ_KQ_L} \,. \label{3chargeentropy_general}
\end{eqnarray}
This provides a map between moduli and electric charges, which allows to define the classes of black holes capable of exploring different regions of moduli space.
The entropy \eqref{3chargeentropy_general} is that of a three-charge, macroscopic black hole in 5 dimensions. The quantities $Q_{I}$ denote the number of M2 branes wrapping their respective 2-cycles.

One relevant observation is that the map between moduli space points and black hole charges is continuous only in the classical approximation. In the quantum theory, charge quantization implies that black hole can only fill a discrete set of points in moduli space. However, this set becomes dense in our main regions of interest, which correspond to infinite distance limits, so that they effectively reproduce a continuous interpolation effectively equivalent to ignoring charge quantization (see \cite{Delgado:2022dkz} for a more detailed discussion). We also follow a similar intuition and utilize continuous interpolation in other regimes, such as in the interior of moduli space. Physically, the heuristic argument is that, given a point in moduli space which is associated to the attractor values of a given black hole, vacua with moduli vevs at nearby points are also sensitive to the existence of this black hole, as they can explore it by letting their scalar fields run with mild gradients along the attractor flow until they attain the desired attractor values. It would be interesting to quantify these effects and provide a precise prescription for the continuous map between moduli and charges, possibly along the lines of \cite{Ooguri:2005vr}. Similar comments apply to the relation between black hole charges and moduli in other dimensions in other sections.

\subsection{Minimal 5d Black Holes as UV Probes in Asymptotic Limits}
\label{sec:5d-asymptotic}

In this section we show that the minimal black holes in M-theory on CY3 follow the species scale along the two qualitatively different asymptotic limits in the vector-multiplet moduli space. We will first discuss the different possible limits in section \ref{sec:5d-limits}, and then probe these limits with black holes in section \ref{sec:5d-BH}.

\subsubsection{The Asymptotic Limits \& their Species Scale}
\label{sec:5d-limits}

It has been shown in \cite{Lee:2019wij} that there exist only three infinite distance limits in the vector-multiplet moduli space of M-theory on CY3 (see e.g. section 5.2.1 of \cite{vandeHeisteeg:2023dlw} for a summary). In all three of these limits, the CY3 ${\mathbf X}_6$ admits a fibration structure, where the fiber is shrinking whilst the base blows up to keep the overall volume fixed. One of these limits, arises in elliptically fibered CY3 ${\mathbf X}_6$, and the limit corresponds to a decompactification to 6d. The other two possible limits arise when the CY3 ${\mathbf X}_6$ is a K3 or ${\mathbb T}^4$ fibration, and the limits correspond to (different) emergent strings. In fact, one observes that these two physically different limits are distinguished by the dimensionality of the fiber and of the base. In particular, the decompactification and emergent string limits correspond to a two- and four-dimensional fiber, respectively. As we will see at the end, this fact will be key for the black holes to reproduce the species scale in both types of limits correctly.

As we approach the infinite distance limit the fibration becomes adiabatic. Thus, the volume of the CY3 can be approximated as
\begin{equation} \label{eq:5d-template}
  \mathcal V \sim x^2 Y \, ,
\end{equation}
where $x$ and $Y$ are two moduli parametrizing the volumes of the four- and two-dimensional part of the geometry, regardless of which one is interpreted as the fiber or the base. As is customary, we are considering a homogeneous limit in the four-dimensional part of the CY3. That is, we are taking $Y^i \sim Y$, where $Y^i$ are the moduli controlling the volumes of the two-cycles that yield the leading contribution to the volume of the four-dimensional part of the geometry as it shrinks or blows up. The role of the more general, non-homogeneous case is to continuously connect different types of limits in which different fibration structures become more relevant \cite{Lee:2019wij}. In what follows, we will not enter into these global and model-dependent aspects of asymptotic limits in moduli space. This type of question will be addressed in Section \ref{sec:9d} for models with fewer moduli.

Equation \eqref{eq:5d-template} is valid for the two qualitatively different types of infinite distance limits, which are distinguished by the limit in which it is assumed to hold. For the emergent string limits, the species scale $\Lambda_s$ is given by the emergent string scale, which arises from an M5-brane wrapped on the K3 or ${\mathbb T}^4$ fiber. Hence we have
\begin{equation}\label{eq:emergentlimit}
  x\to 0 \, , \quad Y \sim x^{-2} \to \infty \, , \quad \Lambda_s \sim Y^{-1/2}\,,
\end{equation}
On the other hand, for the decompactification limit, we have a tower of M2-branes wrapped on the elliptic fiber (which are KK modes in a dual type IIB frame). The species scale is given by the 6d Planck scale. Hence we have
\begin{equation} \label{5d_decompactification_limit}
  Y\to 0 \, , \quad x \sim Y^{-1/2} \to \infty \, , \quad \Lambda_s \sim x^{-1/2}.
\end{equation}
We have indicated in both cases the scaling of the species scale. As found in \cite{vandeHeisteeg:2023dlw}, it coincides with the scale at which the Einstein-Hilbert and the curvature-squared term in the EFT action become of the same order.

We now restrict to the vector multiplet moduli space by requiring\footnote{Typically one sets $\mathcal V =1$, but allowing for a general constant value will be more convenient for our purposes.}
\begin{equation}\label{eq:volconst}
  \mathcal V = \text{fixed} \, .
\end{equation}
This can be achieved by writing
\begin{equation}
  x = e^{\mp \frac{1}{\sqrt{3}} \, \Delta} \, , \quad Y = e^{\pm \frac{2}{\sqrt{3}} \, \Delta} \, .
\end{equation}
In this way, the modulus $\Delta$ is canonically normalized. One can check this using \eqref{eq:volconst} and \eqref{eq:metricon}. We also allow for both signs of $\Delta$, so that $\Delta\to \infty$ can describe both emergent string or decompactification limits. They correspond to the upper and lower sign, respectively.

Let us remark that there could be other moduli that enter the expression of the CY volume and that here we are only taking into account the leading contribution that blows up in the infinite distance limit. We are thus restricting ourselves to a one-dimensional locus in moduli space exploring one of these asymptotic limits. Nevertheless, this should suffice to characterize the physics that are relevant in this strict limit: close enough to the boundary of moduli space, only the leading term in the prepotential matters. In what follows, we use this general frame work to find the relevant minimal black holes that probe each of these two different types of limits. 

\subsubsection{Minimal Black Holes in Asymptotic Limits}
\label{sec:5d-BH}

We are now ready to use the 5d attractor mechanism to find the asymptotic properties of the black holes exploring these asymptotic limit. The only relevant charges to explore this limit are $Q_x$, $Q_Y$, associated to the corresponding moduli. The central charge is given by\footnote{This can be obtained from \eqref{eq:Ze} by setting $Q_i \sim Q_Y$ for the charges that appear with the moduli $Y^i$ that satisfy $Y^i \sim Y$ in the limit. Any other modulus that is not going to zero or infinity can be set to a constant by turning off its related charge.}
\begin{equation}
  Z_e = x Q_x + Y Q_Y = Q_x \, e^{\mp \frac{1}{\sqrt{3}} \, \Delta} + Q_Y \, e^{ \pm \frac{2}{\sqrt{3}} \, \Delta} \, .
\end{equation}
This function can be shown to be extremized at
\begin{equation} \label{eq:modulus}
  \Delta = \pm \frac{1}{\sqrt{3}} \log \left( \frac{Q_x}{2 Q_Y} \right) \, .
\end{equation}
Plugging this back into the central charge we obtain
\begin{equation}
  \left. Z_e \right|_h \sim Q_x^{2/3} \, Q_Y^{1/3} \, .
\end{equation}
Thus, the entropy of the black hole is given by
\begin{equation} \label{eq:entropy_3charge}
  S \sim \left( Z_e \right)^{3/2} \sim Q_x \, Q_Y^{1/2}  \, .
\end{equation}
Having these expressions, we can compute the scale that corresponds to the size of the smallest black hole as we approach any of the asymptotic limits. We start with the emergent string limits before moving on to decompactification limits. In both cases we will find that the smallest black holes that probe these limits follow the species scale.

\subsubsection*{Emergent String Limits}

To describe an emergent string limit, we choose the upper sign in \eqref{eq:modulus}. We see that $\Delta\to\infty$ is explored when $Q_x \gg Q_Y$. In this regime, the BH entropy is minimized if we keep $Q_Y$ fixed. That is, we take 
\begin{equation}
  Q_x \to \infty \, , \quad Q_Y = \text{fixed} \, .
\end{equation}
In particular, the smallest entropy is obtained by taking the smallest possible value for $Q_Y$. Regardless of this value, the entropy scales as
\begin{equation}
  S \sim Q_x \, .
\end{equation}
Furthermore, in this limit we have
\begin{equation}
  \Delta \simeq \frac{1}{\sqrt{3}} \log \left( Q_x \right) \, .
\end{equation}
Finally, we combine these two equations to obtain
\begin{equation}
   \Lambda_{\rm BH} \sim S^{-1/3} \sim Q_x^{-1/3} \sim e^{- \frac{1}{\sqrt{3}} \, \Delta } \, .
\end{equation}
This matches with the species scale obtained in eq. \eqref{eq:emergentlimit}. Thus, the smallest black holes that probe this asymptotic limit correctly capture the species scale.  

\subsubsection*{Decompactification Limit}
For the decompactification limit, we choose the lower sign in \eqref{eq:modulus}. In this case, the entropy is minimized while exploring $\Delta\to\infty$ when
\begin{equation}
  Q_Y \to \infty \, , \quad Q_x = \text{fixed} \, .
\end{equation}
This implies
\begin{equation}
  \Delta \simeq \frac{1}{\sqrt{3}} \log \left( Q_Y \right) \, .
\end{equation}
To get the smaller entropy in this limit, we take the smallest possible value for $Q_x$. Regardless of this value, the entropy scales as
\begin{equation}
  S \sim Q_Y^{1/2} \, .
\end{equation}
Combining these two equations we finally get:
\begin{equation}
   \Lambda_{\rm BH} \sim S^{-1/3} \sim Q_Y^{-1/6} \sim e^{- \frac{1}{2\sqrt{3}} \, \Delta } \, .
\end{equation}
Again, comparing with eq. \eqref{5d_decompactification_limit}, we recover the correct exponential parameter for the species scale in a decompactification limit.

\medskip

Together with the previous result, this shows that the minimal BPS black holes follow the species scale for the two physically different types of limit in the vector multiplet moduli space of M-theory on CY3. It is remarkable that the dimensionality of the fiber with respect to that of the base and the attractor mechanism conspire to make this result hold. Before moving on to more complicated setups, we make a slight detour to examine whether or not the matching between the scale of the smallest classical black holes and the species scale persists in the bulk of moduli space. 

\subsection{Minimal 5d Black Holes in the Interior of Moduli Space}
\label{sec:5d-interior}

In this section we consider the above picture, but exploring the interior of moduli space. We are able to carry out this analysis because it is known that in 5d $\mathcal{N}=1$ compactifications of M-theory that the prepotential is exact. We need a precise expression for the prepotential, so we focus on a particular example. Consider the CY3 given by the intersection of a bidegree (4,1) and bidegree (1,1) hypersurfaces in ${\bf P}_4\times{\bf P}_1$, which has Hodge numbers $h_{1,1} = 2$, $h_{2,1} = 86$, first discussed in \cite{Greene:1995hu,Greene:1996dh}, and used in e.g. \cite{vandeHeisteeg:2023dlw} in this context. The prepotential is
\begin{equation}
\frac{1}{3!}C_{IJK} x^I x^J x^K=\frac 56 x^3 + 2x^2Y
\end{equation}
The relation between $x, Y$ and the physical modulus $\Delta$ is
\begin{equation}\label{eq:modulikk}
x=e^{-\Delta/\sqrt{3}}\quad ,\quad Y=-\frac{5}{12}e^{-\Delta/\sqrt{3}}+\frac 12 e^{2\Delta/\sqrt{3}}
\end{equation}
The K\"ahler cone where this prepotential is valid (namely $x > 0$, $Y > 0$) corresponds to
 \begin{equation}
e^{\sqrt{3}\Delta}>\frac 56 \; \Rightarrow \; \Delta >-0.105263
\end{equation}
The infinite distance point is at $\Delta\to\infty$. Following \cite{vandeHeisteeg:2023dlw}, we take the species scale in the interior of the moduli space to be given by 
\begin{equation} \label{eq:interior-5d}
\Lambda_{s}=\left( \frac {50x+24Y}{12}\right)^{-1/2}=\left(\frac{10}3 e^{-\frac{\Delta}{\sqrt{3}}}+e^{\frac{2\Delta}{\sqrt{3}}}\right)^{-1/2} \, .
\end{equation}
This corresponds to the scale at which the Einstein-Hilbert term and the curvature-squared correction appearing in the effective action ---and whose Wilson coefficient is proportional to the second Chern class of the CY3--- become of the same order.

Now let us consider the size $l_{BH}$ of the classical black holes. The central charge \eqref{eq:Ze} is 
\begin{equation}
Z_e=\frac{1}{12}e^{-\frac{\Delta}{\sqrt{3}}}\left(\left( 6e^{\sqrt{3}\Delta}-5\right) Q_Y+12Q_x\right) \, .
\end{equation}
Using \eqref{eq:modulikk}, we are able to minimize the central charge with respect to the physical modulus $\Delta$. We thus find that the modulus at the horizon and black hole is given by
\begin{equation}
\Delta|_h=\frac{1}{\sqrt{3}}\log \left( \frac{Q_x}{Q_Y}-\frac {5}{12}\right)\quad ,\quad l_{BH}=\frac 12 \left(\frac 32\right)^{\frac 16} Q_Y^{\frac 16}\,\big(\,12Q_x-5Q_Y\,\big)^{\frac 13} \, .
\end{equation}
For the attractor value of the modulus to lie in the K\"aher cone, we need $Q_x/Q_Y>5/4$. 

\begin{figure}[htb]
\begin{center}
\includegraphics[scale=.9]{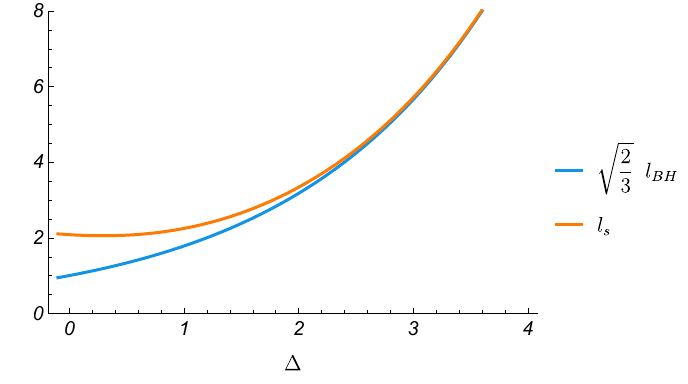} 
\caption{\small The species scale $l_{s} = \Lambda_s^{-1}$ compared to the size of the smallest possible classical black holes $l_{BH}$ (re-scaled by the factor in equation \eqref{eq:rescale}). The results show that the size of the smallest possible classical black holes agrees with the species scale up to at most ${\cal O}(1)$ factor, even in the deep interior of moduli space.   } 
\label{fig:match-5d-interior}
\end{center}
\end{figure}

We now pick the charges such that $l_{BH}$ is the smallest. It is easy to check that this is achieved by setting $Q_Y = 1$, so we can focus on the family parametrized by  $Q_x$, for which
\begin{equation}
l_{BH}=\frac 12 \left(\frac 32\right)^{\frac 16}\,\big(\,12Q_x-5\,\big)^{\frac 13}
\end{equation}
Given that scales are intrinsically defined up to order 1 factors, we match this length with the species scale in \eqref{eq:interior-5d} in the asymptotic region in order to efficiently compare them in the interior of moduli space. One can in fact show analytically that the required rescaling factor is given by
\begin{equation}\label{eq:rescale}
{\rm lim}_{\Delta\to \infty}\left(\frac{\Lambda_{s}^{-1}}{l_{BH}}\right)\to \sqrt{\frac 23}
\end{equation}
Plotting these two scales, we obtain Figure \ref{fig:match-5d-interior}.

The match in the asymptotics is very essentially perfect. This is remarkable, and indicates that the 2-derivative theory contains information about the species scale, as explained in section \ref{sec:5d-asymptotic}. In addition, the match in the interior of moduli space is also very precise. In order to look at this in detail, we can zoom into the deep interior of moduli space, in the region around the desert point $\Delta_{des}=\log(\frac 53)/\sqrt{3}\sim 0.3$, at which the species scale length is shortest. 

This example hence illustrates the surprising fact that these attractors seem to know about the species scale without any UV input, not just asymptotically near infinite distance limits, but even in the interior of moduli space. Given this remarkable feat of the five-dimensional case, we turn to study this in other setups.

\section{4d $\mathcal{N}=2$ from type II String Theory on CY3}
\label{sec:4d}

We now turn to 4d $\mathcal{N}=2$ Calabi-Yau compactifications of type II string theories. The three types of asymptotic limits in vector multiplet moduli space were classified in \cite{Corvilain:2018lgw,Lee:2019wij}. We show that, in contrast with the 5d case, the minimal BPS black holes reproduce the species scale with precision in certain classes of infinite distance limits, while they do not in others. We provide a heuristic understanding of this behavior, which moreover suggests interesting links with the microscopic picture. For a class in which the asymptotic behavior gives the correct species scale, we also explore the match in the interior of moduli space, and find a semi-quantitative match (i.e. up to ${\cal O}(1)$ factors) similar to the 5d case.

\subsection{Structure of the 4d Theory and Black Holes}
\label{sec:4d-background}

In this section we review the structure of the 4d $\mathcal{N}=2$ supergravity theory and of the black holes we consider. We phrase the discussion in the language of type IIA compactification on CY3 (a similar discussion may be carried out in terms of type IIB theory), and we follow conventions in e.g. \cite{Cribiori:2022nke} (see e.g. appendix A in \cite{Cribiori:2022cho} for a quick review).

We consider 4d $\mathcal{N} = 2$ supergravity arising from type IIA string theory compactified on a Calabi-Yau threefold $\mathbf{X}_6$. We focus on the $h_{1,1}+1$ gauge bosons, and the $h_{1,1}$ complex scalars in vector multiplets. The latter are expressed in terms of projective coordinate fields $x^I = (x^0, x^i)$, with $i=1, \ldots, h_{1,1}$. The 4d effective action is specified by a prepotential, which in the large volume limit has the form
\begin{equation}
F(x^I)=\frac 16 C_{ijk} \frac{x^i x^j x^k}{x^0}
\label{general-prep}
\end{equation}
where $C_{ijk}$ are the triple intersection numbers, just like in the 5d case in section \ref{sec:5d-background}. The $x^I$ together with $F_I = \partial_I F$ form holomorphic symplectic sections $(x^I,F_J)$ and parametrize the moduli of a symplectic basis of even-dimensional cycles $\{ A^I, B_J\}$ on $\mathbf{X}_6$.  In terms of the $\{x^I,F_J\}$, the Kähler potential is given by:
\begin{equation}\label{eq:Kahlerpotdef}
    \mathcal{K}=-\log \; i \left( \bar x ^I F_ I - x^I \bar F _ I  \right)\,.
\end{equation} The physical vector multiplet moduli are given by
\begin{equation}
z^{i}=\frac{x^i}{x^0} \, .
\end{equation}
As it is customary, we will split these complex moduli into an axion and a saxion
\begin{equation}\label{eq:tomoduli}
    z^i = \phi^i + i t^i \, .
\end{equation}
The metric on moduli space is given by the following formula: 
\begin{equation}
    g_{i\bar{j}}dz^i d\bar{z}^ j = 2 \, \partial _ i \bar{\partial _ j} \mathcal{K} \;dz^i d\bar{z}^ j\,.
\end{equation}
With this in hand, we now review the classification of infinite distance limits in the moduli space of these theories. 

\subsubsection{Classification of Infinite Distance Limits}
\label{sec:4d-limits}

While remaining in the large volume regime, infinite distance limits in the vector multiplet moduli space of these theories correspond to sending a subset of the saxions to infinity, i.e. $t^a \sim \tl \to \infty$ with $a=1,\ldots,\mathbf{n}$ and $\mathbf{n}\leq h_{1,1}$, while all the axions are kept fixed. In general, the other saxions can be kept fixed at very large values ---so as to stay in the large volume regime--- or sent to infinity at slower rates. The second possibility allows for a continuous interpolation between asymptotic regimes (see e.g. \cite{Castellano:2023jjt}). As in Section \ref{sec:5d-limits}, we do not wish to address this type of global and model-dependent aspects of infinite-distance limits in moduli space here. Thus, we will focus on simple limits in which the saxions other than $t^a$ are kept fixed. In Section \ref{sec:9d}, we address how to analyze the global structure of infinite-distance limits in a more tractable setup.

These infinite-distance limits have been carefully classified in \cite{Grimm:2018ohb,Corvilain:2018lgw,Lee:2019wij} by different means (see \cite{Castellano:2023jjt} for a nice summary). Following the notation in \cite{Grimm:2018ohb}, we will denote them as Type IV, III and II. Before entering into the details of each type of limit, let us recall that the 4d dilaton is in a hypermultiplet. Thus, since we will only explore the vector multiplet moduli space, it must be kept fixed. This imposes the co-scaling (see e.g. \cite{Lee:2019wij})
\begin{equation} \label{eq:co-scaling}
    g_s^2 \sim \mathcal V = \frac{1}{6} C_{ijk} t^i t^j t^k \, ,
\end{equation}
where $g_s$ is the string coupling and $\mathcal V$ is the CY volume in string units. For later convenience, we note that this is equivalent to imposing $M_s \sim M_{Pl,4}$, where $M_s$ and $M_{Pl,4}$ are the string and the 4d Planck scales, respectively. Let us analyze the different limits in turn.

\paragraph{Type IV: Decompactification to five dimensions} \mbox{ } \\
Geometrically, this type of limit arises when the CY volume blows up uniformly. That is, we have $t^a \sim \tl \to \infty$ such that
\begin{equation}
    \mathcal V \sim \tl^3 \, .
\end{equation}
Similarly, the prepotential in \eqref{general-prep} is also \emph{cubic} in the  moduli whose saxions are being sent to infinity. The leading tower in this limit can be shown to be composed by D0-branes. Hence, the limit corresponds to a decompactification to the 5d theory obtained by compactifying M-theory on the same CY. Crucially taking \eqref{eq:co-scaling} into account, the mass scale of this KK tower satisfies
\begin{equation}
    \frac{(M_{\rm KK})_{\rm IV}}{M_{Pl,4}} \sim g_s^{-1} \frac{M_s}{M_{Pl,4}} \sim \tl^{-3/2} \, .
    \label{kk-scale-iv-limit}
\end{equation}
Consequently, the species scale is given by the 5d Planck scale, i.e., $\Lambda_s \sim M_{Pl,5}$. As usual, this scale is related to the 4d Planck scale by
\begin{equation}\label{species5d}
    \Lambda_s^3 \, (M_{\rm KK})_{\rm IV}^{-1} \sim M_{Pl,4}^{2} \, ,
\end{equation}
such that the species scale can be found to scale with $\tl^{-1/2}$ in 4d Planck units.

\paragraph{Type III: Decompactification to six dimensions} \mbox{ } \\
Type III limits are allowed when the CY admits is elliptic fibered. In particular, the infinite distance limit corresponds to blowing up the four-dimensional base while the volume of the elliptic fiber remains constant in string units. Hence, the CY volume satisfies
\begin{equation}
    \mathcal V \sim \tl^2 \, .
\end{equation}
In the same way, we see that the prepotential in \eqref{general-prep} is \emph{quadratic} in the moduli whose saxions are being sent to infinity. The leading tower is composed by D0-branes and D2-branes wrapping the elliptic fiber, such that this limit corresponds to a decompactification to the 6d theory obtained by compactifying F-theory on the same CY. The mass scale of this KK tower is given by
\begin{equation}
    \frac{(M_{\rm KK})_{\rm III}}{M_{Pl,4}} \sim g_s^{-1} \frac{M_s}{M_{Pl,4}} \tl^{-1}\, ,
    \label{kk-scale-iii-limit}
\end{equation}
where we have taken \eqref{eq:co-scaling} into account. Furthermore, the species scale corresponds to the 6d Planck scale  and thus is determined by
\begin{equation}
    \Lambda_s^4 \, (M_{\rm KK})_{\rm III}^{-2} \sim M_{Pl,4}^{2} \, .
\end{equation}
As for the previous case, we recover that the species scale goes as $\tl^{-1/2}$ in 4d Planck units.

\paragraph{Type II: Emergent string limits} \mbox{ } \\
Finally, Type II limits can happen when the CY admits a K3 or an Abelian (i.e. ${\mathbb T}^4$) fibration. In this case, the infinite distance limit is reached when the volume of the two-dimensional base blows up while that of the fibration remains fixed in string units. That is,
\begin{equation}
    \mathcal V \sim \tl \, .
\end{equation}
Something similar holds for the prepotential in \eqref{general-prep}, which is found to be \emph{linear} in the modulus whose saxion is being sent to infinity. In this limit, the lightest object is given by an NS5-brane wrapping the fiber, thus yielding an emergent string limit.\footnote{Even though it will not be of relevance for us, let us note that a K3 or Abelian fibration leads to a heterotic or type IIB emergent string, respectively.} Therefore, both the species scale and the mass scale of the excitation modes of the emergent string coincide asymptotically. This scale satisfies
\begin{equation} \label{eq:scale-TypeII}
    \frac{\Lambda_s}{M_{Pl,4}} \sim \frac{\sqrt{T_{\rm NS5}}}{M_{Pl,4}} \sim g_s^{-1} \mathcal V_{(4)}^{1/2} \frac{M_s}{M_{Pl,4}} \sim \tl^{-1/2} \, ,
\end{equation}
where we have taken into account \eqref{eq:co-scaling} and that the volume of the 4d fiber in string units ---denoted by $\mathcal{V}_{(4)}$--- remains constant in this limit. Incidentally, this coincides with the mass scale of various particle towers, as they are required to guarantee that the emergent string is ten-dimensional. In particular, we will later use that the mass scale of D0-branes, D2-branes wrapped 2-cycle within the 4d fiber, and D4-branes wrapping the 4d fiber does coincide with that in \eqref{eq:scale-TypeII}. Indeed, as they are only sensitive to volumes within the fiber ---which remain constant in string units in the limit, all these mass scales are suppressed by $g_s^{-1}\sim \tl^{-1/2}$. 

\medskip

As one can observe, the different types of limits are nicely distinguished from the EFT perspective by having a cubic, quadratic or linear behavior of the prepotential with the moduli whose saxions are being sent to infinity. For our purposes, it will be useful to consider a single model that captures all types of infinite distance limits. We will thus consider the so-called STU model, that corresponds to taking a CY with three K\"ahler moduli and $C_{(123)}=1$ as the only non-vanishing triple intersection numbers. This leads to
\begin{equation}\label{eq:ourprep}
    F(x^I)=\frac{x^1 x^2 x^3}{x^0} \, .
\end{equation}
In the standard notation for the STU model, one would replace $x^1$, $x^2$ and $x^3$ by $S$, $T$ and $U$. Given this prepotential, we see that Type IV, III and II limits correspond to sending three, two or one of the saxions to infinity at the same rate.

\medskip

Before moving on, let us discuss how these different UV scales show up in higher-curvature corrections appearing in the low-energy EFT action. As shown in \cite{vandeHeisteeg:2022btw,Cribiori:2022nke,vandeHeisteeg:2023ubh,Calderon-Infante:2023uhz, Castellano:2023aum, vandeHeisteeg:2023dlw, Bedroya:2024ubj, Aoufia:2024awo,Castellano:2024bna,Calderon-Infante:2025ldq}, the species scale precisely shows up suppressing the curvature-squared corrections with respect to the Einstein-Hilbert term. The Wilson coefficient of this higher-curvature correction is given by the genus-one topological string free energy, $F_1$, such that $F_{1} \sim \Lambda_s^{-2}$. For any asymptotic limit, $F_1$ behaves linearly with $\tl$, which nicely explains why the species scale satisfies
\begin{equation}\label{eq:species}
    \frac{\Lambda_s}{M_{Pl,4}} \sim \tl^{-1/2} \, ,
\end{equation}
regardless of the type of limit that is being explored. Additionally, the KK scale in Type III and IV limits can be seen to appear suppressing further higher-dimensional operators whose Wilson coefficients are given by higher-genus topological string free energy, $F_g$ with $g>1$. More concretely, one finds that this operator of dimension $n=2g+2$ is suppressed by the KK scale to the usual field-theoretic power, i.e., $F_g \sim M_{\rm KK}^{d-n}$ where $d=4$ is the number of spacetime dimensions \cite{Castellano:2023aum}. The interplay between this two types of scales and how they suppress different higher-dimensional operator has been recently explored in \cite{Calderon-Infante:2025ldq}.

\subsubsection{BPS Black Hole Probes}

We now consider BPS black holes in this theory, carrying electric and magnetic charges under the $h_{1,1}+1$ $U(1)$ gauge fields. In the realization of these BPS black holes in the type IIA on CY3 picture, the electric charges correspond to $q_0$ D0-branes at a point in $\mathbf{X}_6$, $q_i$ D2-branes wrapped on the $h_{1,1}$ 2-cycles, while the magnetic ones correspond to $p^i$ D4-branes on the dual 4-cycles and $p^0$ D6-branes wrapped on the whole $\mathbf{X}_6$. Clearly, the notion of electric or magnetic depends on a choice of polarization, and so can change upon symplectic transformations.

The central charge of these black holes is
\begin{equation}\label{eq:Z}
Z =  e^{\mathcal{K}/2}(p^I F_ I-q_I x^I ) \, .
\end{equation}
Such 4d black holes have an attractor mechanism which fixes the values of the vector multiplet moduli at the horizon (first constructed in \cite{Ferrara:1995ih} and further studied in \cite{Ferrara_1996a,Ferrara_1996b,Ferrara_1997,moore,Denef_1999,Denef:2001xn}). The attractor equations are 
\begin{equation}\label{eq:4dn2attractor}
p^I=\text{Re}[C x^I]_h\quad ,\quad q_J=\text{Re}[C F_J]_h \, ,
\end{equation}
with $C= 2 i e^{\mathcal{K}/2} \bar{Z}$ and the ``$h$" subscript indicates that these quantities are evaluated at the horizon. The entropy is given by
\begin{equation} \label{eq:entropy}
S=\pi |Z|^2_h \, .
\end{equation}
In what follows, the ``$h$" subscripts are implicit. 

\medskip

Let us now turn to the STU model introduced around \eqref{eq:ourprep}. The general solution to the attractor equations in this model was found in \cite{LopesCardoso:1996yk}.\footnote{Their notation for the charges is related to ours by setting: $M_0=-q_0$, $M_1=-p_1$, $M_2=q_2$, $M_3=q_3$, $N_0=p_0$, $N_1=q_1$, $N_2=p_2$, $N_3=p_3$.} We wish to explore infinite distance at the horizon of these black holes by taking some limit for the black hole charges. This limit should be such that the axions go to some constant value.\footnote{Black hole families with varying axion values are on general grounds expected to explore the same limits, hence leading to the same results as those considered in our work.} Given that the particular value of the axion does not change the properties of the infinite distance limit, we fix $\phi^i =0$ for simplicity. For solutions to the attractor equations in \cite{LopesCardoso:1996yk}, this condition becomes
\begin{equation} \label{eq:vanishing-axions-condition}
    q_0 p^0 = q_1 p^1 = q_2 p^2 = q_3 p^3 \, .
\end{equation}
One can distinguish seventeen ways of solving for this condition. Sixteen of them come from choosing different combinations of vanishing charges such that \eqref{eq:vanishing-axions-condition} vanishes (for each of the four values of $I$, there is a two-fold choice of setting either $q_I$ or $p^I$ to zero). The other way of solving is by imposing that \eqref{eq:vanishing-axions-condition} should be non-vanishing. Notice that the last solution nicely interpolates between all the others, in the sense that any of the former options can be recovered from the latter by setting some of the charges to zero.\footnote{In fact, this implies that some of the sixteen solutions making \eqref{eq:vanishing-axions-condition} vanish can be grouped together with that in which \eqref{eq:vanishing-axions-condition} is non-vanishing. At the end of the day, there are fifteen independent solutions.}

Even though there are many ways of setting the axions to zero, we are interested in regular-sized black holes at the classical level. For this, the solution to the attractor equation should sit at a regular point in the moduli space and the central charge should be non-vanishing. In the model at hand, the first condition means that $t^1$, $t^2$ and $t^3$ should take some positive and finite value. As explained in \cite{LopesCardoso:1996yk}, this implies that
\begin{equation}
    D \equiv (q_0 p^1 + q_2 q_3) (p^0 q_1 + p^2 p^3) - ( q_0 p^0 + q_1 p^1 - q_2 p^2 - q_3 p^3)^2 > 0 \, .
\end{equation}
Using \eqref{eq:vanishing-axions-condition}, this yields
\begin{equation}
    (q_0 p^1 + q_2 q_3) (p^0 q_1 + p^2 p^3) > 0 \, .
\end{equation}
In particular, notice that neither of the two quantities in parenthesis can vanish. For instance, this implies that if $q_0 = 0$ then we must have $q_2, q_3 \neq 0$. Taking into account this type of constraints, only two out of the sixteen ways of making \eqref{eq:vanishing-axions-condition} vanish survives. They correspond to setting $p^0 = q_1 = q_2 = q_3 = 0$ or $q_0 = p^1 = p^2 = p^3 = 0$ while the other charges must be non-vanishing. Together with the option of having all the charges to be non-vanishing, we end up with the three families of BHs that we describe in more detail next.

\paragraph{D0/D4-brane black holes:} The only non-vanishing charges for the first family of black holes are $q_0$, $p^1$, $p^2$ and $p^3$. The attractor point is given by
\begin{equation} \label{ttt_D0D4D4D4}
    t^1 = \sqrt{\frac{q_0p^1}{p^2p^3}} \,, \qquad
    t^2 = \sqrt{\frac{q_0p^2}{p^1p^3}} \,, \qquad
    t^3 = \sqrt{\frac{q_0p^3}{p^1p^2}} \,,
\end{equation}
and the entropy of the black hole is
\begin{equation} \label{ent_D0D4D4D4}
S=2\pi \sqrt{ q_0 p^1p^2 p^3  } \,.
\end{equation}

\paragraph{D2/D6-brane black holes:} In this case, the non-vanishing charges are $q_1$, $q_2$, $q_3$ and $p^0$. The attractor point is given by
\begin{equation} \label{ttt_D6D2D2D2}
    t^1 = \sqrt{\frac{q_2q_3}{p^0q_1}} \,, \qquad
    t^2 = \sqrt{\frac{q_1q_3}{p^0q_2}} \,, \qquad
    t^3 = \sqrt{\frac{q_1q_2}{p^0q_3}} \,,
\end{equation}
while the black hole entropy is
\begin{equation} \label{ent_D6D2D2D2}
S=2\pi \sqrt{  p^0 q_1 q_2 q_3  } \,.
\end{equation}

\paragraph{D0/D2/D4/D6-brane black holes:} There is a third family of black holes, which belongs to this general class, but whose particular features have not been highligthed in the literature. All the charges are non-vanishing for this family of black holes, subject to the restriction (c.f. \eqref{eq:vanishing-axions-condition})
\begin{equation} \label{qipi_all_equal}
    q_0 p^0 = q_1 p^1 = q_2 p^2 =q_3 p^3 \neq 0 \, .
\end{equation}
There are two especially convenient ways of parametrizing this solution, namely by using $\left\{q_0,p^0,p^1,p^2,p^3\right\}$ or $\left\{q_0,q_1,q_2,q_3,p^0\right\}$ as the independent charges. Using these two parametrizations, the attractor point is given by
\begin{equation} \label{ttt_D0D2D4D6}
    t^1 = \sqrt{\frac{q_0p^1}{p^2p^3}} = \sqrt{\frac{q_2q_3}{p^0q_1}} \,, \qquad
    t^2 = \sqrt{\frac{q_0p^2}{p^1p^3}} = \sqrt{\frac{q_1q_3}{p^0q_2}} \,, \qquad
    t^3 = \sqrt{\frac{q_0p^3}{p^1p^2}} = \sqrt{\frac{q_1q_2}{p^0q_3}} \,,
\end{equation}
while the entropy reads
\begin{equation} \label{ent_D0D2D4D6}
    S = 2\pi \sqrt{\frac{q_0}{p^1 p^2 p^3}} \left| p^1 p^2 p^3 + q_0 (p^0)^2 \right| = 2\pi \sqrt{\frac{p^0}{q_1 q_2 q_3}} \left| q_1 q_2 q_3 + (q_0)^2 p^0 \right| \, .
\end{equation}
These two parametrizations allow the formulas \eqref{ttt_D0D2D4D6} and \eqref{ent_D0D2D4D6} to capture the D0/D4 and the D2/D6 cases discussed above, when \eqref{qipi_all_equal} is not satisfied: The first parametrization would recover the D0/D4 black holes upon setting $p^0=0$, while the second would recover the D2/D6 system for $q_0=0$. The existence of a general formula for all families does not mean that the the D0/D4, D2/D6 and D0/D2/D4/D6 families will behave in the same way when exploring the asymptotic limits.
As will become clear in later sections, the behavior of the minimal black holes through \eqref{ent_D0D2D4D6} may be controlled by a leading term which disappears when some of the charges is set to zero. This underlies the fact that the results obtained for D0/D4 or D2/D6 black holes may differ from those obtained through the generic entropy formula \eqref{ent_D0D2D4D6}.

To better understand this family of solutions, we may consider realizing it in a ${\mathbb T}^6$ compactification, which for simplicity we take to be factorized as $(\mathbb T^2)^3$. Performing three T-dualities, one on each 2-torus, the different branes map to D3-branes wrapped on 3-cycles. In this picture it is easy to realize that the constraints (\ref{qipi_all_equal}) imply that the whole  D0/D2/D4/D6 system is mapped to {\em a single} stack of $N$ D3-branes wrapped on a factorizable 3-cycle $\Pi$ in $(\mathbb T^2)^3$ (see \cite{Rabadan:2001mt} for a similar argument). Indeed, define a basis $\{[a_i],[b_i]\}$ of 1-cycles in the $i^{th}$ $\mathbb T^2$, and consider the 3-cycle $[\Pi]=\otimes_i (n_i[a_i]+m_i[b_i])$, with ${\rm gcd}(n_i,m_i)=1$ for each $i$. The charges are obtained by decomposing it in the basis of 3-cycles (given by the products of 1-cycles), and are given by
\footnote{To get the identification \eqref{charges_vs_wrapping_nb}, we perform three T-dualities along the $[a_i]$.}
\begin{eqnarray} \label{charges_vs_wrapping_nb}
&& p^0=N\, n_1 n_2 n_3\, , \quad p^1=N\, n_1 m_2 m_3 \, , \quad p^2=N\, m_1 n_2 m_3 \, , \quad p^3=N\, m_1 m_2 n_3 \nonumber\\
&& q_0=N\, m_1 m_2 m_3\, , \quad q_1=N\, m_1 n_2 n_3 \, , \quad q_2=N\, n_1 m_2 n_3 \, , \quad q_3=N\, n_1 n_2 m_3 
\end{eqnarray}
which automatically satisfy (\ref{qipi_all_equal}). This means that, in this toroidal setup, the original D0/D2/D4/D6 systems is a single stack of D6-branes carrying constant worldvolume magnetic fluxes (i.e. magnetized as in \cite{Angelantonj:2000hi}), which produce lower-dimensional induced D0/D2/D4 charges satisfying the constraint. These so-called magnetized branes, and the T-dual branes on 3-cycles are familiar from the model building literature with magnetized and/or intersecting D-brane models, see \cite{Blumenhagen:2000fp,Aldazabal:2000dg,Aldazabal:2000cn,Blumenhagen:2000ea}, also \cite{Blumenhagen:2006ci,Ibanez:2012zz} for reviews. In the context of black holes, D-branes with this kind of charge constrains have been studied in  \cite{Eckardt:2023nmn,Bena:2022wpl,Bena:2024oeq,Dulac:2025toappear}.

The above explanation implies that in the toroidal setup, the D-brane system preserves 16 supersymmetries.\footnote{The general case can be addressed by performing one T-duality, upon which the $p^I$ and $q_I$ charges are mapped to a D1-D5-D3-D3 system with a common direction and orthogonal D3-D3 branes. The condition \eqref{qipi_all_equal} translates into $N_1N_5=N_{3,I}N_{3,II}$, which is the condition for having 16 supersymmetries \cite{Eckardt:2023nmn}.} In a generic CY, the system preserves less supersymmetry. The above enhancement becomes only manifest in the large volume limit of the type IIA side, in which the D6-brane worldvolume fluxes are dilute. Alternatively, by performing mirror symmetry as three T-dualities we reach the type IIB side in the large complex structure limit, in which the 3-cycles wrapped by the D3-brane align and mimic a single stack.

Using \eqref{charges_vs_wrapping_nb}, one can rewrite the entropy \eqref{ent_D0D2D4D6} as
\begin{equation} \label{ent_D0D2D4D6_with_nm}
S=2\pi N^2 \sqrt{(n_1n_2n_3)(m_1m_2m_3)} \, |n_1n_2n_3+m_1m_2m_3| = 2\pi \sqrt{q_0p^0} \, |q_0+p^0| \,.
\end{equation}
This formula is however only valid for toroidal compactifications and with a particular choice of factorization of the 3-cycle $[\Pi]$. Hence, for our general discussion we use the general CY formula \eqref{ent_D0D2D4D6}. 

\medskip

For each of the three different types of black holes, the attractor point provides a map between black hole charges and moduli space. Regarding the issue of charge quantization making black hole attractors explore only a discrete (but possibly dense) subset of moduli space, we take the same viewpoint expressed at the end of section \ref{sec:5d-background}, and simply proceed implicitly using a smooth interpolation to make the map continuous.

Having described the asymptotic limits in the moduli space of 4d $\mathcal{N}=2$ and the BPS black hole solutions that probe moduli space, we now move on to our main goal: identifying and characterizing the smallest black holes that probe the three different types of limits.

\subsection{Minimal 4d Black Holes as UV Probes in Asymptotic Limits}
\label{sec:4d-limits-BH}

In this section, we explore infinite distance limits with the attractor mechanism by considering families of black holes in which some of the charges are sent to infinity, as already mentioned.  We proceed by considering Type IV, III and II limits separately. We will consider the STU model as a convenient template, and exploit the above three families of black holes to explore each of these limits. The strategy is to determine, within each family of black holes, the smallest ones exploring each limit, defined by some subset of the saxions diverging with an overall parameter $\tl$ as $t^a\sim \tl\to \infty$. 

We will see that, very remarkably, even though the black holes are built from the 2-derivative EFT action, the corresponding minimal black hole scale turns out to have a non-trivial UV interpretation. In particular, the size of such minimal black holes corresponds to UV scales such as the mass scale of KK towers or the species scale $\Lambda_{s}$. We also provide heuristic arguments clarifying the microscopic reasons explaining which scale is reproduced in each case. The key idea is the comparison of the sets of diverging charges of the black hole family exploring the limit with the sets of states in the corresponding asymptotic towers. The intuition is that the diverging charges, which enter the entropy formula and produce the UV scale, are related to the exponentially large number of ways the black hole can be dressed by new states from some tower becoming light in the limit. 

\subsubsection{Type IV: Decompactification to Five Dimensions}
\label{sec:4d-limits-BH-iv}

As already anticipated, this corresponds to the limit in which the CY blows up uniformly. In the STU model, this limit corresponds to taking $t^1 \sim t^2 \sim t^3 \sim \tl \to \infty$. Note that, even though there is one single effective independent modulus, we maintain the individual labels for the different charges for clarity. Let us now consider the three different families of black holes in turn.

\paragraph{D0/D4-brane black holes:} \mbox{ } \\
From \eqref{ttt_D0D4D4D4}, we see that for this family the black hole horizon explores this limit when
\begin{equation}
    q_0 \gg p^1 \sim p^2 \sim p^3 \, .
\end{equation}
The entropy in \eqref{ent_D0D4D4D4} is minimized if we further take $p^1 \sim p^2 \sim p^3 \sim \mathcal O(1)$, while $q_0\to\infty$. In this case, \eqref{ttt_D0D4D4D4} yields $\tl \sim \sqrt{q_0}$ and the black hole entropy reads
\begin{equation} \label{limit_typeIV_D0/D4}
    S = 2 \pi \sqrt{q_0 p^1p^2p^3} \sim \sqrt{q_0} \sim \tl \, .
\end{equation}
Comparing to \eqref{species5d} or \eqref{eq:species}, we see that this corresponds to a species scale sized black hole in this decompactification limit to 5d. Thus, we find that the smallest black hole within this family precisely probes this UV scale in the asymptotic limit. Just like in the 5d case, this is a remarkable outcome, since the derivation from the classical black hole does not involve any knowledge of the UV structure of the theory. 

Notice that the charge configuration that we have considered nicely fits with the expectation for a minimal black hole probing the species scale in this limit. Indeed, we are blowing up the D0-brane charge ---which corresponds to the leading tower in this limit--- while the others remain of order one. This limit has already been discussed in e.g. \cite{Cribiori:2022nke,Cribiori:2023ffn,Calderon-Infante:2023uhz} 

\paragraph{D2/D6-brane black holes:} \mbox{ } \\
In this case, using \eqref{ttt_D6D2D2D2}, the $t^1 \sim t^2 \sim t^3 \sim \tl \to \infty$ limit is explored by taking
\begin{equation}
    q_1 \sim q_2 \sim q_3 \gg p^0 \, .
\end{equation}
Furthermore, within the family of black holes with this scaling of charges, the entropy \eqref{ent_D6D2D2D2} is minimized when $p^0$ is kept fixed, while $q_1 \sim q_2 \sim q_3 \sim q \to \infty$. Equation \eqref{ttt_D6D2D2D2} then yields $\tl \sim \sqrt{q}$, while the entropy of this family of black holes is given by
\begin{equation}
S = 2 \pi \sqrt{p^0 q_1q_2q_3} \sim q^{3/2} \sim \tl^3 \sim \mathcal{V} \,.
\end{equation}
We thus recover a family of black holes probing the KK scale in the decompactification to 5d limit (c.f. \eqref{kk-scale-iv-limit}). Again, we find that the family of smallest classical black holes within this family explores a UV scale in this limit. 

In this case, the fact that the black hole reproduces the KK scales also admits a microscopic explanation. Since we have an order one D6-brane charge, in the 5d M-theory frame the system corresponds to a black hole placed at the center of a Taub-NUT geometry (see e.g. \cite{Dijkgraaf:2006um}). As discussed in \cite{Castellano:2025ljk}, this explains geometrically why the size of the circle at the horizon coincides with the radius of the horizon. 

\paragraph{D0/D2/D4/D6-brane black holes:} \mbox{} \\
Finally, let us consider the third family of black holes. Using the $\left\{q_0,p^0,p^1,p^2,p^3\right\}$ parametrization, we see that Type IV limit is explored when 
\begin{equation} \label{typeIV_pq_hierarchy}
    q_0 \gg p^1 \sim p^2 \sim p^3 \, ,
\end{equation}
similar to the D0/D4-brane black holes. Given the constraint in \eqref{qipi_all_equal}, this can be achieved with $q_0 \sim q_i \gg p^0 \sim p^i = \mathrm{fixed}$, or with $q_0 \gg p^i, q_i \gg p^0 = \mathrm{fixed}$. The entropy of this family of black holes is minimized while exploring the limit in the first hierarchy case.\footnote{This can be seen by plugging $p^1 \sim p^2 \sim p^3 \sim p$ and $\sqrt{q_0} \sim \sqrt{p} \, \tl$ with $\tl\to\infty$ into \eqref{ent_D0D2D4D6}. This makes the two terms in \eqref{ent_D0D2D4D6} manifestly quadratic in $p$ and $p^0 t$, respectively. Thus, the entropy is minimized when $p$ and $p^0$ are kept fixed in the $\tl\to\infty$ limit.} For $p^0 \neq 0$, the leading term in \eqref{ent_D0D2D4D6} leads to
\begin{equation}
    S = 2\pi \sqrt{\frac{q_0}{p^1 p^2 p^3}} (p^0)^2 \left|q_0 \right| \sim q_0^{3/2} \sim \tl^3 \, ,
\end{equation}
thus leading to a KK sized black hole. This leading term vanishes when $p^0 =0$, which leads to the D0/D4 black hole that probes the species scale as above. Thus, we see how the general formula \eqref{ent_D0D2D4D6} nicely encode the two possible sizes for the smallest black holes in this limit, namely the KK or the species scale. 

Microscopically, one can again interpret the non-vanishing D6-brane charge as putting the black hole on a Taub-NUT geometry in 5d M-theory, thus leading to the identification of the black hole size with the KK scale. When turning off this charge, we end up with a D0/D4 black hole that is allowed to probe smaller lengths, in this case yielding the species scale.

\medskip

Above we have discussed several specific families of minimal BPS black holes in the STU model, and explored the UV scales probed by the horizon of its smallest representatives. In particular we have found that the sizes of such minimal black holes are always equal or larger than the species scale. This suggests that perhaps a general argument can be made to prove such a bound in full generality. In fact, in section \ref{sec:proofIV}, we will argue that indeed there cannot be such a family of BPS black holes becoming parametrically smaller than the species scale asymptotically.

\subsubsection{Type III: Decompactification to Six Dimensions}
\label{sec:4d-limits-BH-iii}

We now move on to the Type III limit, which in the STU model corresponds to taking e.g. $t^2 \sim t^3 \sim \tl \to \infty$ while $t^1$ remains fixed. As in the previous limit, we study the three families of black holes in turn.

\paragraph{D0/D4-brane black holes:} \mbox{ } \\
From \eqref{ttt_D0D4D4D4}, we see that the black hole horizon explore this limit when
\begin{equation}
    p^2 \sim p^3 \, , \quad q_0 \, p^1 \sim p^2 p^3 \, , \quad q_0 \gg p^1 \, .
\end{equation}
Imposing this, the entropy in \eqref{ent_D0D4D4D4} is minimized if we keep $p^1 \sim \mathcal O(1)$ while $q_0 \to \infty$. Equation \eqref{ttt_D0D4D4D4} then yields $\tl \sim \sqrt{q_0}$ and the black hole entropy reads
\begin{equation}
    S = 2 \pi \sqrt{q_0 p^1p^2p^3} \sim q_0 \sim \tl^2 \, .
\end{equation}
As discussed in \eqref{kk-scale-iii-limit}, this corresponds to a black hole of the size of the extra dimensions in the decompactification limit to 6d, so it remarkably corresponds to a UV, albeit a KK scale rather than the species scale. 

The fact that classical black holes reproduce a UV scale is still remarkable. Note also that their failure to reproduce the species scale is to be expected. The leading tower in this limit is given by D0/D2-brane bound states, so a black hole built out of these species and exploring the asymptotic limit is expected to have $q_0 \sim q_1 \to \infty$, with other charges fixed (or subleading in the limit). On the other hand, the black holes that we have just described have a diverging $q_0 \sim p^1 p^2 \to \infty$, but $q_1$ is actually vanishing (furthermore, the charges $p^1$ and $p^2$ are also blowing up, although in a subleading way). Hence they do not have the appropriate composition to be specifically sensitive to the tower producing the species scale.

\paragraph{D2/D6-brane black holes:} \mbox{ } \\
This family of black holes explore the Type III limit when 
\begin{equation}
    q_2 \sim q_3 \, , \quad p^0 \, q_1 \sim q_2 q_3 \, , \quad q_1\gg p^0  \, .
\end{equation}
The entropy in \eqref{ent_D0D4D4D4} is minimized if we further keep $p^0 \sim \mathcal O(1)$ while $q_1 \to \infty$. Equation \eqref{ttt_D0D4D4D4} then yields $\tl \sim \sqrt{q_1}$ and the black hole entropy reads
\begin{equation}
    S = 2 \pi \sqrt{p^0 q_1q_2q_3} \sim q_1 \sim \tl^2 \, .
\end{equation}
Thus, this family of black holes also recovers the KK scale in Type III limits. Again, not recovering the species scale is to be expected. In this case, we have $q_1 \to \infty$ as expected, but $q_0$ is vanishing.

\paragraph{D0/D2/D4/D6-brane black holes:} \mbox{ } \\
Finally, let us consider the third family of black holes. Using the $\left\{q_0,p^0,p^1,p^2,p^3\right\}$ parametrization, we see that Type III limit is explored when 
\begin{equation}
    p^2 \sim p^3 \, , \quad q_0 \, p^1 \sim p^2 p^3 \, , \quad q_0 \gg p^1 \, .
\end{equation}
as for the D0/D4-brane black holes. In this case, the entropy is minimized by fixing $p^0$ and $p^1$ while $q_0\to\infty$.\footnote{This can be seen by substituting $p^2 p^3 \sim q_0 \, p^1$ and $\sqrt{q_0} \sim \tl \sqrt{p^1}$ into \eqref{ent_D0D2D4D6}. This makes the two terms in \eqref{ent_D0D2D4D6} manifestly quadratic in $q_0$ and $p^0$, such that the entropy is minimized when both of them are kept fixed as $\tl$ blows up.} Taking all this into account, \eqref{ttt_D0D2D4D6} yields $\tl \sim \sqrt{q_0}$ and \eqref{ent_D0D2D4D6} leads to
\begin{equation}
    S \sim q_0 \sim \tl^2 \, .
\end{equation}
We again recover a family of black holes of the size of the KK scale in the Type II limit. Interestingly, we have $q_0 \sim q_1 \to \infty$ as expected for a family of black holes following the species scale asymptotically. Nevertheless, these are not the only charges that are blowing up. Indeed, we also have $q_2 \sim q_3 \sim p^2 \sim p^3 \sim \sqrt{q_0} \to \infty$. Even though they are blowing up at a lower rate, these extra charges provide a significant contribution to the black hole entropy, thus explaining why these black holes do not reproduce the species scale.

\medskip

We have seen that the three families of black holes recover the KK scale when exploring this limit. In contrast with Type IV limits, we do not find a family of black holes that follow the species scale. One may again wonder whether this is true in general. Using similar methods as in the other types of limit, in this case we will argue in section \ref{sec:proofIII} that there can indeed be no family of BPS black holes that satisfy the attractor equations \eqref{eq:4dn2attractor} and have an entropy that scales as 
\begin{equation}
    S \sim |z|^a\,, \text{ with } a<2 \; \text{ as } |z|\to \infty\,.
\end{equation}
This implies that there are no families of BPS black holes that follow the species scale asymptotically.

\paragraph{4d vs 5d Black holes in $\mathbb{T}^2$ decompactification limits} \mbox{ } \\[-.3cm]

Our result that the different families of 4d minimal black holes do not reproduce the species scale in the Type III limit may seem puzzling, given that 5d black holes managed to reproduce it in all possible infinite distance limits. Indeed, in the 4d setup the Type III limits arise for elliptically fibered CY compactifications with the volume of the base growing faster than that of the fiber. This is essentially the same limit as in the $\mathbb{T}^2$ decompactification limit in 5d \eqref{5d_decompactification_limit}, which also probes the same types of $\mathbb{T}^2$ fibration for the Calabi-Yau space, with the base growing and the fiber shrinking in size. It is therefore striking that such limit in 5d admits extremal black holes that follow the species scale, which it does not in 4d. We aim to provide a rationale for this difference in what follows.

In the 5d setup, the attractor mechanism stabilization of the moduli at the black hole horizon results from a competition between M2-branes wrapping various 2-cycles of the CY. Using the notation of section \ref{sec:5d-BH}, the number $Q_Y$ denotes the M2-brane charge wrapping the fiber, and $Q_{x_i}$ the charges of M2-branes wrapping the 2-cycles in the base. For simplicity, we can just consider two charges $Q_{x_1}= Q_{x_2}\equiv Q_x$ on the base. In order to follow the species scale, the M2-brane charges should scale like
\begin{equation}
Q_Y \sim \lambda^{a_\mathrm{fiber}} \,, \qquad Q_{x_1} = Q_{x_2} \sim \lambda^{a_\mathrm{base}} \,,
\end{equation}
where $a_\mathrm{fiber}$ and $a_\mathrm{base}$ are positive integer numbers that cannot vanish simultaneously.
The sizes of the 2-cycles are stabilized according to
\begin{equation}
x^I=\frac{\mathcal{V}}{x_I} = \frac{(Q_x Q_x Q_Y)^{1/3}}{Q_I} \,.
\end{equation}
So, demanding that the volume of the base grows (equivalently, that the volume of the fiber decreases) yields
\begin{equation}
    a_\mathrm{fiber} > a_\mathrm{base} \,.
\end{equation}
Besides, demanding that the entropy \eqref{eq:entropy_3charge} follows the species scale ($S\sim |t| = Y$) implies 
\begin{equation}
a_\mathrm{fiber}+2 a_\mathrm{base}=1 \,.  
\end{equation}
This admits a solution, $a_\mathrm{fiber}=1$, $a_\mathrm{base}=0$, which is realized by the black hole family in section \ref{sec:5d-BH}.

On the other hand, in the 4d setup, there is one additional modulus that needs to be stabilized: the radius of the M-theory circle, or in type IIA string theory, the ten dimensional dilaton. Indeed, taking the four dimensional dilaton to be constant (it is in a hypermultiplet) imposes that one should co-scale the CY volume and the ten dimensional dilaton as in \eqref{eq:co-scaling}. Therefore, also from the M-theory perspective, there is an extra variable to stabilize. Morally, the stabilization of the additional variable (volume of the fiber) is taken care of by the presence of a fourth black hole charge: the type IIA D6-brane charge $p^0$. The brane charges scale as
\begin{equation}
q_1 \sim \lambda^{a_\mathrm{fiber}} \,, \qquad q_2 \sim q_3 \sim \lambda^{a_\mathrm{base}} \,, \qquad p^0 \sim \lambda^{a_\mathrm{D6}} \,,
\end{equation}
with the coefficients $a_i$ being positive integers that are not all vanishing.
Using \eqref{ttt_D6D2D2D2}, requiring that the volume of the base grows faster than the volume of the fiber implies that 
\begin{equation}
a_\mathrm{D6} + 2 \, a_\mathrm{base} > 3 \, a_\mathrm{fiber} \,,
\end{equation}
and asking that the CY volume increases implies that
\begin{equation}
3 \, a_\mathrm{D6} > a_\mathrm{fiber} + 2 \, a_\mathrm{base} \,.
\end{equation}
However, wishing that the scaling follows the species scale ($S\sim |t|$) requires
\begin{equation}
a_\mathrm{D6} + a_\mathrm{base} =0 \,,
\end{equation}
which cannot be satisfied.

\subsubsection{Type II: Emergent string limits}
\label{sec:4d-limits-BH-ii}

Consider now the Type II limit, which in the STU model corresponds to e.g. $t^1 \sim\tl\to \infty$ while $t^2 \sim t^3$ remains fixed. As usual, we study the three families of black holes in turn.

\paragraph{D0/D4-brane black holes:} \mbox{ } \\
From \eqref{ttt_D0D4D4D4}, we see that the Type II limit corresponds to
\begin{equation}
    q_0 \sim p^1 \gg p^2 \sim p^3 \, .
\end{equation}
Imposing this, the entropy in \eqref{ent_D0D4D4D4} is minimized for $p^2 \sim p^3 \sim \mathcal O(1)$ while $q_0 \sim p^1 \to \infty$. Equation \eqref{ttt_D0D4D4D4} then yields $\tl \sim q_0$ and the black hole entropy reads
\begin{equation} \label{entropy_typeII_D0D4D4D4}
    S = 2 \pi \sqrt{q_0 p^1p^2p^3} \sim \sqrt{q_0p^1} \sim q_0 \sim \tl \, .
\end{equation}
As discussed in \eqref{eq:scale-TypeII} and \eqref{eq:species}, this recovers the species scale. Since the leading tower is that of the excitations of a string, this also coincides with the scale of the tower asymptotically. 

This example is surprising, not only in that the classical black holes reproduce a UV scale, but also that it corresponds to the species scale of an emergent string limit. Note that in this case none of the charges of the black hole is related to the string excitation modes, which are neutral under the gauge fields that enter the attractor mechanism. Hence it would seem that the black holes manage to reconstruct the species scale of a tower, despite not having the appropriate composition to be sensitive to it. The microscopic explanation for this behaviour is the following. Let us recall that the leading string tower is accompanied by several other (less dense) KK towers, whose role is to reconstruct six extra dimensions and make the emergent string ten-dimensional. In fact, both $q_0$ and $p^1$ are related to these KK towers, that become massless at the same rate as the emergent string. Indeed, as discussed after \eqref{eq:scale-TypeII}, the mass scales of D0-branes and D4-branes wrapping the fiber in the Type II limit are of the same order as that of the emergent string. In this way, this family of black holes naturally explores the species scale, albeit in an indirect way.

\paragraph{D2/D6-brane black holes:} \mbox{ } \\
For this family of black holes, the Type II limit is explored when 
\begin{equation}
    q_2 \sim q_3 \gg p^0 \sim q_1 \, .
\end{equation}
The entropy in \eqref{ent_D0D4D4D4} is minimized if we further keep $p^0 \sim q_1 \sim \mathcal O(1)$ while $q_2 \sim q_3 \to \infty$. Equation \eqref{ttt_D0D4D4D4} then yields $\tl \sim q_1$ and the black hole entropy reads
\begin{equation}
    S = 2 \pi \sqrt{p^0 q_1q_2q_3} \sim q_1 \sim \tl \, .
\end{equation}
This family of black holes also recovers the species scale in Type II limits. This coincides with the mass scales of D2-branes wrapping 2-cycles within the fiber. Thus, as in the previous case, the dominant charges $q_2$ and $q_3$ correspond to states within the KK towers accompanying the emergent string. 

\paragraph{D0/D2/D4/D6-brane black holes:} \mbox{ } \\
Finally, let us consider the third family of black holes. Using the $\left\{q_0,p^0,p^1,p^2,p^3\right\}$ parametrization, we see that Type II limit is explored when 
\begin{equation}
    q_0 \sim p^1 \gg p^2 \sim p^3 \, .
\end{equation}
as for the D0/D4-brane black holes. In this case, the entropy is minimized by fixing $p^0$, $p^2$ and $p^3$, while $q_0 \sim p^1 \to\infty$.\footnote{This can be seen by plugging $q_0 \sim p^1$, $p^2 \sim p^3$ and $q_0 \sim p^1 \sim \tl p^2 \sim \tl p^3$ into \eqref{ent_D0D2D4D6}. This makes the two terms in \eqref{ent_D0D2D4D6} manifestly quadratic in $p^0$ and $p^2$, such that the entropy is minimized when both of them are kept fixed as $\tl$ blows up.} Then \eqref{ttt_D0D2D4D6} yields $\tl \sim q_0$ and \eqref{ent_D0D2D4D6} leads to
\begin{equation}
    S \sim q_0 \sim \tl \, .
\end{equation}
We again recover a family of black holes of the size of the species scale in the Type II limit. The leading black hole charges are $q_0 \sim q_2 \sim q_3 \sim p^1 \to \infty$, while the others remain fixed. As in previous cases, these dominant charges correspond to KK towers accompanying the emergent string.

\medskip

As in the previous cases, it is natural to wonder whether there are BPS black holes that are parametrically smaller than the species length scale asymptotically. In section \ref{sec:proofII}, we use similar techniques as in the previous cases to argue that this is indeed not possible.

\medskip

In conclusion, we have shown that 4d minimal classical black holes, despite requiring only EFT ingredients for their construction, can probe UV scales of the theory in which they are embedded, in the different infinite distance limits. We have moreover provided heuristic arguments explaining in which cases the probed scale is the species scale, or an independent KK scale. The argument interestingly involves a comparison of the divergent charges of the black hole and the states in the asymptotic towers in the limit (more on this in section \ref{sec:micro}). Before that, we would like to explore the behaviour of minimal 4d black holes in the interior of moduli space, to which we turn next.

\subsection{A Lower Bound on the Entropy}
\label{sec:lowerbound}
So far we have discussed several specific families of BPS black holes in the STU model, and explored the UV scales probed by the horizon of its smallest representatives. In particular, we have found that the sizes of such minimal black holes are always equal or larger than the species scale. This suggests that perhaps a general argument can be made to prove such a bound in full generality. In fact, in the following we argue that indeed there cannot be such black holes that are parametrically smaller than the species scale asymptotically. 

In this section, we perform our analysis for a general Calabi-Yau compactification, extending our results from the STU model in the previous sections. The difference between the two setups is the possibility for subleading terms in the prepotential, potentially involving extra spectator moduli. For simplicity, we focus on single-moduli limits, in which a single saxion blows up as $t^1 \sim \tl \to \infty$. In what follows, we argue that there can be no family of black holes that satisfy the attractor equations \eqref{eq:4dn2attractor} and have an entropy that scales as \begin{equation}\label{eq:whatwewannaprove}
    S \sim \tl^n\,, \text{ with } n<1 \; \text{ as } \tl \sim t^1 \to \infty\,.
\end{equation}
We will make no extra assumptions on the CY prepotential in \eqref{general-prep} but, as stressed above, we focus on limits in which only the saxion of the modulus $z^1$ blows up while the other saxions remain fixed at finite (but large) values. Depending on the triple intersection numbers of the from $C_{1jk}$, the leading term in the prepotential goes as $(z^1)^a$ with $a\in \{1,2,3\}$, corresponding to Type II, III and IV limits \cite{Corvilain:2018lgw}. Since we have at least one physical modulus, we have at least 4 charges: $\{p^0,q_0\}$ corresponding to the graviphoton and $\{p^1,q_1\}$ corresponding to the U(1) associated to $z^1$. For each extra spectator modulus, we will have another pair of charges.

The general formula for the entropy is given by \eqref{eq:entropy}
\begin{equation}
    S= \pi e^{\mathcal{K}} |p^I F_I - q_ I x^I|^2 \,,
\end{equation}
where $I \in \{0,1,2,3,\cdots\}$. Let us re-write this formula in terms of the physical moduli $z^I= \{1, x^i/x^0\}$ \eqref{eq:tomoduli}, with $i \in \{1,2,3\cdots\}$. It is known that the B-periods $F_I$ are holomorphic in the $X^I$ with degree 1 which implies that $F_J= F_J(z) \,x^0$, where the B-periods $F_J(z)$ are now expressed in terms of the physical moduli $z^i$. This and \eqref{eq:Kahlerpotdef} imply that the formula for the entropy remains unchanged under this reparametrization.\footnote{Indeed, the transformation $x^I \to z^I x^0$ can be understood as a K\"ahler transformation $\mathcal{K}\rightarrow \mathcal{K}- f -\bar f \;\;\; C\rightarrow e^{-f} C \;\;\; x^I \rightarrow e^{f} x^I \;\;\; F \rightarrow e^{2f} F(x^I)$ which leaves the charges of the black hole unchanged (see for instance \cite{Ooguri:2004zv}).} That is, we have
\begin{equation}\label{eq:Sproof}
   S=\pi e^{\mathcal{K}(z)} |p^I F_I(z) - q_ I z^I|^2\,,
\end{equation}
where neither $\mathcal{K}(z)$ nor $F_I(z)$ depend on $x^0$. Notice though that the charges depend on $x^0$ and $C$ through the attractor equations \eqref{eq:4dn2attractor}, which we can write as: 
\begin{align}
p^J&=\text{Re}[C x^0 z^J]_h = \frac{1}{2} [ C x^ 0 z^J + \bar C \bar x^0 \bar z ^ J]_h\,,\\
q_J&=\text{Re}[C x^0 F_J(z)]_h = \frac{1}{2} [ C x^ 0 F_J(z) + \bar C \bar x^0 \bar F_J(z)]_h\,.
\end{align}
Plugging this into the entropy \eqref{eq:Sproof}, we find:
\begin{equation}
    S=  \frac12 \pi| C x^0 |^2 |\text{Im}[\bar z^I  F_I (z)]|\,.
\end{equation}
In what follows, we will use this expression to argue that there are no non-trivial solutions that scale as \eqref{eq:whatwewannaprove}.
The reason is that, as we will see, no matter how the leading term in the prepotential scales with $\tl$ as $\tl \to \infty$, $|\text{Im}[\bar z^I  F_I (z)]|$ will grow as a high enough power of $\tl$ to make it difficult to engineer solutions that follow \eqref{eq:whatwewannaprove}. In fact, as we will now show, it is even difficult to find solutions that are small enough to follow the species scale as $\tl \to \infty$. We will even be able to prove that this is impossible for a Type II limit, as we observed in section \ref{sec:4d-limits-BH-iii}. As we will argue below, we expect that although probing the species scale was possible for certain limits in the STU model, as we add spectator moduli in subleading terms in the prepotential, this will get increasingly difficult and we expect generic solutions to be a lot larger than the species scale as $\tl \to \infty$.

We now consider each of the three different cases separately, for clarity. In each case we show that there can be no black holes that become parametrically smaller than the species scale, and comment on the implications of having (or not) subleading terms. In the Type III case we can in fact argue that there cannot be solutions that are smaller or equal to the species scale, as observed in section \ref{sec:4d-limits-BH-iii}.

\subsubsection{Type IV: $a=3$}\label{sec:proofIV}
In this case the prepotential has a leading term that is cubic in $z^1$. We can write it schematically as 
\begin{equation}\label{eq:FtypeIV}
    F(x^I)= C_{111} \frac{(x^1)^3}{x^0}+ \cdots \, ,
\end{equation}
where the ellipsis denotes possible subleading terms in the asymptotic limit including other spectator moduli. Then, using \eqref{eq:tomoduli} and \eqref{eq:Sproof}, one finds that the entropy scales as: 
\begin{equation} \label{entropy-tl3}
  S =  2 \pi| C x^0 |^2  C_{111} \tl^3 + \cdots \, .
\end{equation}

Therefore, the condition \eqref{eq:whatwewannaprove} can be translated into the condition that $|C x^0|^2$ scales like $\tl^{-b}$ with $b>2$ as $ \tl \to \infty$. The contradiction arises when we consider how the integer quantized charges of this black hole are related to this quantity, through the attractor equations themselves \eqref{eq:4dn2attractor}. To see this, let us consider the charges $p^0$ and $p^1$:
\begin{equation}
    p^0 = \text{Re}[C x^0]\,, \;\; p^1 = \text{Re}[C x^1]=  \text{Re}[C x^0 z] \,.
\end{equation}
We see immediately that if $|C x^0|^2\sim\tl^{-b}$ with $b>2$ at leading order in $\tl$ as $  \tl \to \infty$, then $p^0$ and $p^1$ must vanish from the start, as they cannot take values arbitrarily close to zero because of Dirac quantization. This has the consequence of trivializing the whole solution, as one can show explicitly that if both $p^0$ and $p^1$ vanish, then $q_0$ and $q_1$ must vanish as well. Indeed, imposing $p^0=0$ and $p^1=0$ comes down to solving the following system of equations:
\begin{align}
    p^0 &= \text{Re}[C x^0] = 0 \label{eq:p0}
    \\
    p^1 &= \phi^1 \,\text{Re}[C x^0] - \tl \, \text{Im}[C x^0] =0
\label{eq:p1}
\end{align}
The only solutions to this system lead to vanishing $C$ and/or $x^0$ which trivializes the solution completely as can be seen from \eqref{eq:entropy} and \eqref{eq:4dn2attractor}.

We have thus shown that there cannot be black holes that are parametrically smaller than the species scale as $\tl \to \infty$. Let us note however that this does not guarantee that there is a solution that follows the species scale. In section \ref{sec:4d-limits-BH-iv}, we saw that there is such a solution for the STU model: this solution can be recovered by picking $|C x^0|^2\sim\tl^{-2}$ above. However, for a general Calabi-Yau there are other terms in the prepotential \eqref{eq:FtypeIV}. For example, if we add an extra modulus $z^2$ in the prepotential as follows:
\begin{equation}\label{eq:FtypeIV2}
    F(x^I)= C_{111} \frac{(x^1)^3}{x^0}+ C_{222} \frac{(x^2)^3}{x^0} \, ,
\end{equation}
then all of the discussion above still applies. If we allow $t^2$ to blow up at the same rate as $t^1$ at the horizon, we would find ourselves in a case similar to that of section \ref{sec:4d-limits-BH-iv}, where we know there are black holes that follow the species scale. However, if we treat $z^2$ as a spectator modulus, whose value should be large (to stay in the large volume/large complex structure limit) but \textbf{finite}, then, satisfying $p^2 = \text{Re}[C x^0 z^2]$ implies that $p^2$ should vanish if we want the solution to follow the species scale. Indeed, we should have $|C x^0|^2\sim\tl^{-2}$, and so we see that $p^2$ has to be set to zero, due to charge quantization. This leads to the added condition: 
\begin{equation} \label{eq:spectator-charge}
    p^2 = \phi^2 \,p^0 - t^2 \, \text{Im}[C x^0]=0
\end{equation}
This, in combination with \eqref{eq:p0} leads to vanishing $C$ and/or $x^0$ which trivializes the solution completely. Therefore, the only non-trivial black hole solutions in the presence of this spectator modulus that probe the $t^1 \sim \tl\to \infty$ limit will have to be larger than the species scale. More generally, we expect that the solution will only follow the species scale when all of the moduli are blowing up, which is the case when all magnetic charges are non-vanishing and finite.

Given this result, it is interesting to ask how much larger than the species scale the BH must be. Imposing that either $p^0$ or $p^2$ should not be vanishing ---otherwise the black hole solution trivializes--- leads to the statement that $|C x^0|$ should not go to zero as $\tl \to \infty$. Plugging this into \eqref{entropy-tl3}, we obtain that, in the presence of this $z^2$ spectator modulus,
\begin{equation}
    S \gtrsim \tl^3 \, .
\end{equation}
In other words, the black hole must be at least of the size of the KK scale. It is remarkable that, precisely in the case in which it is not allowed to have species scale sized black holes, we are automatically pushed to the KK scale. 

It would be interesting to understand in more generality in which cases it is possible to build families of black holes following the species scale, some KK scale, or under which conditions they are necessarily larger than both of them.

\subsubsection{Type III: $a=2$} \label{sec:proofIII}
In this case the prepotential has a leading term that is quadratic in $z^1$. We can write it schematically as follows: 
\begin{equation}\label{eq:FtypeIII}
    F(x^I)= C_{112} \frac{(x^1)^2 (x^2)}{x^0} + \cdots \, .
\end{equation}
For this specific case, we can prove using the same methods as above that there can be no family of black holes that satisfy the attractor equations \eqref{eq:4dn2attractor} and have an entropy that scales as 
\begin{equation} \label{eq:condIII}
    S \sim \tl^n\,, \text{ with } n<\mathbf{2} \; \text{ as } \tl \to \infty\,.
\end{equation}
This implies that there are no families of black holes that follow the species scale asymptotically. This provides an explanation for why the three families of black holes discussed in section \ref{sec:4d-limits-BH-iii} recover the KK scale when exploring this limit. We now detail the proof, using similar methods as above. 
Using \eqref{eq:FtypeIII}, \eqref{eq:tomoduli} and \eqref{eq:Sproof}, one finds that the entropy scales as: 
\begin{equation}
  S =  2 \pi| C x^0 |^2  C_{112} \tl^2\, t^2\,.
\end{equation}

Therefore, equation \eqref{eq:condIII} can be translated into the condition that $|C x^0|^2$ scales like $\tl^{-b}$ with $b>0$ as $\tl \to \infty$. Indeed, as previously, $t^2$ is a spectator modulus which has to remain large but finite, so it cannot depend on $\tl$. We once more show that this forces all the charges to vanish and the solution to trivialize. To see this, let us consider the charges $p^0$ and $p^2$:
\begin{equation}
   p^0 = \text{Re}[C x^0]\,, \;\; p^2 = \text{Re}[C x^2]=  \text{Re}[C x^0 z^2] \,.
\end{equation}
We see immediately that if $|C x^0|^2\sim\tl^{-b}$ with $b>0$ at leading order in $\tl$ as $\tl \to \infty$, then $p^0$ and $p^2$ must vanish from the start, as they cannot take values arbitrarily close to zero because of Dirac quantization. As in the previous case, this has the consequence of trivializing the whole solution.

Again, it would be interesting to better understand the role of adding subleading corrections to the prepotential. Although one would generally expect this to make it harder to find black hole solutions that are small enough to follow UV scales (as explained in the previous section), one could hope that by fine tuning the subleading term, one might be able to engineer a solution that follows the species scale. 

\subsubsection{Type II: $a=1$}\label{sec:proofII}
In this case the prepotential has a leading term that is linear in $z^1$. For simplicity, let us take the case in which it looks like 
\begin{equation}
    F(x^I)= C_{123} \frac{(x^1)(x^2)(x^3)}{x^0}+ \cdots \, .
\end{equation}

Following the same reasoning as in the Type IV case, we can show that there can be no family of black holes that become parametrically smaller than the species scale as $\tl \to \infty$  as in \eqref{eq:whatwewannaprove}. In the limit we are considering, only the saxion of $z^1$ blows up, whilst the other moduli remain finite. Using \eqref{eq:entropy} and \eqref{eq:Z}, we obtain
\begin{equation}
    S= 2\pi |C x^0|^2\,C_{123} \tl \,t^2 t^3\, .
\end{equation}
Therefore, the condition \eqref{eq:whatwewannaprove} translates into $|C x^0|^2$ scaling like $\tl^{-b}$ with $b>0$ as $\tl \to \infty$. Indeed, to stay in the large volume regime, the values of $t^2$ and $t^3$ have to be large but finite, so they cannot depend on $\tl$. As before, this causes all the charges to vanish and the solution to completely trivialize. To see this, let us consider the charges $p^0$, $p^2$ and $p^3$:
\begin{equation}
    p^0 = \text{Re}[C x^0]\,, \;\; p^2 = \text{Re}[C x^2]=  \text{Re}[C x^0 z^2] \,, \;\;p^3 = \text{Re}[C x^3]=  \text{Re}[C x^0 z^3]\,.
\end{equation}
We see immediately that if $|C x^0|^2\sim\tl^{-b}$ with $b>0$ at leading order in $\tl$ as $  \tl \to \infty$, then $p^0$, $p^2$ and $p^3$ must vanish from the start, as they cannot take values arbitrarily close to zero because of Dirac quantization. As in the Type IV case, one can show that imposing these conditions forces the whole solution to trivialize. 

\subsection{Minimal Black Holes in the Interior of Moduli Space in 4d $\mathcal{N}=4$}
\label{sec:4d-interior}

In this section we carry out a similar comparison as above, but in the interior of moduli space. In order to do this, we need to consider concrete models for which the species scale is known even in the bulk of moduli space.  We thus focus on the illustrative example of type IIB string theory on K3$\times\mathbf{T}^2$ \footnote{Clearly, the same examples can be described in terms of dual pictures, e.g. heterotic on $\mathbf{T}^6$, with a mere reinterpretation of charges and moduli.} (or $\mathbf{T}^4\times\mathbf{T}^2$) in a Type II limit of the complex structure modulus of the $\mathbf{T}^2$. The idea is then to compare the size of minimal black holes in this theory with the species scale given by the higher curvature terms through the one-loop topological free energy $F_1$ \cite{vandeHeisteeg:2022btw}. This example has enhanced supersymmetry, which facilitates the analysis. It would be interesting to extend the exploration to genuine 4d ${\cal N}=2$ examples. 

The higher curvature definition of the species scale is given by $\Lambda_s \sim 1/\sqrt{F_1}$, in terms of the topological 1-loop free energy $F_1$ \cite{vandeHeisteeg:2022btw}. For the complex structure modulus $\tau$ of $\mathbf{T}^2$ in type IIB on K3$\times\mathbf{T}^2$, one has \cite{Ferrara:1991uz,Bershadsky:1993ta}:
\begin{equation}\label{eq:dedeking}
  F_1(\tau) = - \log \left( \tau_2 |\eta^2(\tau)|^2 \right) \, ,
\end{equation}
where $\eta$ is the Dedekind eta function.  

BPS black holes in this compactification have been extensively studied in the literature, and we refer the reader to \cite{Dabholkar:2012zz} for a review and references. In our present discussion we follow the conventions and results of this review, sketching the main points relevant for our computation of the black hole species scale in the interior of moduli space.

We consider black holes with vectors of quantized electric and magnetic charges $P$, $Q$. The attractor value for the modulus $\tau =  a + i s $ 
\begin{equation}
  s = \frac{\sqrt{Q^2 P^2 - (Q\cdot P)^2}}{P^2} \, , \quad a = \frac{Q \cdot P}{P^2} \, ,
\end{equation}
We focus on the $a=0$ case, which suffices for the class of black holes we are going to study, hence we set $Q \cdot P = 0$. Under this simplification, we can regard both $P$ and $Q$ as 28-dimensional vectors of magnetic/electric integer charges under the 28 different U(1) gauge fields in the theory.

\begin{figure}[!tbp]
  \centering
  \includegraphics[width=0.8\textwidth]{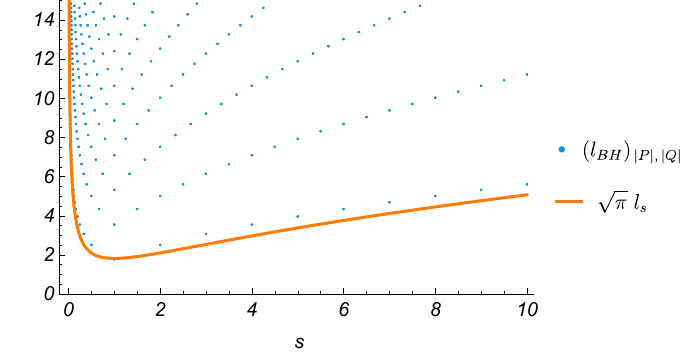}\label{fig:f2}
  \caption{The sizes of classical black holes are shown in blue dots, whereas the species scale as given by equation \eqref{eq:dedeking} is shown in orange. The species scale and the size of the minimal (smallest) black holes match in the bulk of moduli space up to $\mathcal{O}(1)$ factors. The normalization is chosen such that both scale agree at the desert point $s=1$.  }
  \label{figure:4dn4}
\end{figure}

The black hole entropy is given by
\begin{equation}
  S_{BH} = \pi \sqrt{Q^2 P^2 - (P \cdot Q)^2} = \pi  |Q| |P| \, ,
\end{equation}
where in the last step we have used $Q\cdot P =0$. Our goal is to compare the size of these black holes with the species scale given by \eqref{eq:dedeking}. We can now compare the size of the black holes to that the species scale, defined respectively as:
\begin{equation} \label{eq:F1ll}
  l_{s}^{(F_1)} \sim \sqrt{F_1} \, , \quad l_{(BH)} \sim \sqrt{S_{BH}} \, .
\end{equation}
As in the 5d case, in order to fix ${\cal O}(1)$ factors in the comparison, we match the two quantities at a point, which we choose to be the desert point, $s=1$. In figure \ref{figure:4dn4}, we plot the size of black holes as a function of the attractor point $s$, for different values of $|P|$ and $|Q|$.\footnote{For simplicity, we take $|P|$ and $|Q|$ to be integers. Even though, in general, the modulus of a vector of integers is not an integer, the extra possible values for $|P|$ and $|Q|$ do not change our conclusions concerning minimal BHs.} The smallest of such black holes are the minimal black holes. We also show the species scale as given by \eqref{eq:F1ll} as a function of $s$. From Figure \ref{figure:4dn4},  we learn that the two scales agree in the bulk of moduli space up to $\mathcal{O}(1)$ factors. This provides another example where the naive scale obtained by the minimal black holes gives a good estimate of the species scale, even deep in the bulk of moduli space!

In contrast with the 5d case, the match in the asymptotics is not exact here due to the logarithmic corrections to the species scale as given by $F_1(\tau)$, as was already noted in \cite{Cribiori:2023sch} (see also \cite{Calderon-Infante:2023uhz} for related discussions). Indeed, these minimal black holes have a size that goes linearly with $s$ as $s\to \infty$, whilst $F_1(\tau)$ in \eqref{eq:dedeking} blows up schematically as $s+ \log s$. In this way, the black holes only follow the species scale at leading order. It is tantalizing to propose that the additional terms account for corrections to black holes beyond the classical approximation. We leave this interesting direction for future work. 

As a last remark, we would like to mention that the above black hole computation remains equally valid if we replace the K3 by a $\mathbf{T}^4$, hence reproducing the species scale in this case. This is in contrast with the evaluation of the species scale using the $F_1$ higher derivative correction, which vanishes due to the flatness of $\mathbf{T}^4$, i.e. to the additional supersymmetry (see e.g. \cite{Kiritsis:1997gu} for review and references). It is really remarkable that classical black holes can probe the species scale in this setup in which other prescriptions do not apply.

\section{
Black holes in Bottom-up Examples: a 9d Case Study}
\label{sec:9d}

In this section we change perspective and consider classical minimal black holes and their associated scales in bottom-up models, to study the interplay with the existence of an actual UV completion. The motivation is to initiate the exploration of our proposal that sizes of minimal black holes are always larger than or equal to the species length as a swampland criterion to be obeyed by any consistent EFT. 

For this purpose, we focus on a particular example, based on a bottom-up 9d theory, which may be regarded as a truncation of 9d maximal supergravity, in which axions have been removed {\em ad hoc}. The theory provides a simple arena with only two different real moduli and three types of charges which allows us to be fairly exhaustive about the infinite-distance limits that are probed. In particular, we will be able to derive a convex hull formulation capturing the asymptotic exponential rate of $\Lambda_{\rm BH}$ in any such infinite-distance limit in Section \ref{sec:global picture}. The relation with 9d supergravity and hence to circle compactifications of type II string theories allows to propose putative species scales in these limits, and their comparison with the bottom-up black hole scales.

\subsection{Structure of the 9d Theory and Black Holes}
\label{sec:class-9DBHs}

We start with the bosonic sector of the 9d maximal supergravity action. The effective action can be obtained from compactifying type II supergravity actions on a circle \cite{Bergshoeff:1995as,Das:1996je}. In what follows, we will adopt the conventions of \cite{Das:1996je}, and use type IIA language to describe the field content. We focus on the part of the action that involves gravity, the moduli and the 1-form gauge fields, in the string frame it can be written as: 
\begin{equation}\label{eq:fullact}
\begin{split}	
	S_{9d} = \frac{1}{2} \int d^9 x \sqrt{-g} \, &\left[ e^{-2 \phi} \left( R + 4 (\partial \phi)^2 - \frac{1}{4} (\partial \log \chi)^2 - \frac{1}{4} \chi \, (F^{(2)}_2)^2 - \frac{1}{4} \chi^{-1} (F^{(3)}_2)^2 \right) \right. \\
	& \left. - \frac{1}{2} \chi^{-\frac{1}{2}} (\partial a)^2 - \frac{1}{4} \chi^{\frac{1}{2}} \left( F^{(1)}_2 + a F^{(3)}_2 \right)^2  + \cdots\right] \, ,
\end{split}
\end{equation}
where the ellipses are meant to indicate the presence of other terms in the action that play no role in our discussion. There are two saxions, the 9d dilaton $\phi$ and the string frame radion $\chi$, as well as an axion $a$.

The gauge field labeled by the index $(1)$ corresponds to the RR 1-form, the one with index $(2)$ comes from the metric and finally, the one with index $(3)$ comes from dimensionally reducing the kinetic term of the KR 2-form. 

The actual theory we would like to focus on corresponds to a truncation of the above theory, in which we remove the axion by setting it to zero. The resulting theory should be regarded as a bottom-up 9d EFT, whose consistency we plan to explore using classical minimal black holes and their UV scales. The corresponding action in the Einstein frame is:
\begin{equation}
\begin{split}
    S_{9d}= \frac12 \int d^ 9 x \sqrt{-g} &\left[ R - (\partial \hat \phi)^2 - (\partial \hat \chi)^2 - \frac14 e^{2 \hat \chi + \frac{2}{\sqrt{7}}\hat \phi} (F_2^{(2)})^2 \right.  \\ 
    &\left. - \frac14 e^ {- 2\hat \chi + \frac{2}{\sqrt{7}}\hat \phi} (F_2^{(3)})^2  -\frac{1}{4} e^ { \hat \chi - \frac{5}{\sqrt{7}}\hat \phi } (F_2^{(1)})^2 + \cdots \right] \,,
\end{split}
\end{equation}
where the canonically normalized moduli are given by: 
\begin{align}
    \hat \phi = -\frac{2}{\sqrt{7}} \phi\,, \;\;\;\; \hat \chi = \frac12 \log \chi\,.
\end{align}
The action thus reduces to a nine-dimensional Einstein-Maxwell-Dilaton theory. We thus set-out to find extremal charged black hole solutions to this action. We will start by finding a $AdS_2 \times S^7$ near-horizon geometry. We do so by using the entropy function formalism of \cite{Sen:2005wa}. This formalism allows one to determine the value of the scalars at the horizon and the entropy as a function of the charges for extremal black holes. The existence of a near-horizon geometry does not in general guarantee the existence of a full black hole solution (that extends all the way to asymptotic flat space) \cite{Goldstein:2005hq}. For this reason, after discussing the near horizon geometry, we will provide a simple realization of the full black hole solution. 

\paragraph{The near-horizon geometry} \mbox{} \\
Following \cite{Sen:2005wa}, we choose an extremal near-horizon ansatz of the form
\begin{equation}
\begin{split} 
      ds^{2} &= \frac{v}{\beta} \left( -r^2 dt^2 + \frac{1}{r^2} dr^2 \right) + v \, d\Omega_{7}^{2} \, , \\
	  F_{rt}^{(1)} &= e_1 \, , \quad F_{rt}^{(2)} = e_2 \, , \quad F_{rt}^{(3)} = e_3 \, ,\\
	  \partial_\mu s &= 0 \, , \quad \partial_\mu \chi = 0 \, .
\end{split}
\end{equation}
The entropy function in \cite{Sen:2005wa} is defined by
\begin{equation} 
  \mathcal E (Q_i,e_i,s,\chi,v,\beta) = 2\pi \left(  \sum_{i=1}^{3} e_i Q_i - \int_{S^7} d\Omega_7 \sqrt{-g} \, \left.\mathcal L\right|_{h} \right) \, .
\end{equation}
where the subindex $h$ indicated evaluating the Lagrangian on the near horizon ansatz above. We then proceed to the extremisation of the entropy function. Extremizing with respect to the $e_i$, to $\beta$ and $v$, and to $\hat \chi$ and $\hat \phi$ (in that order), we obtain: 
\begin{align}
     \frac{\partial \mathcal E}{\partial e_1} &= 0 \quad \rightarrow \quad e_1 = \frac{3  Q_1 e^{5 \hat \phi/\sqrt{7}- \hat \chi}}{\pi ^4 \beta v^{5/2}}\, , \\
 \frac{\partial \mathcal E}{\partial e_2} &= 0 \quad \rightarrow \quad e_2 =\frac{3  Q_2 e^{-2 \hat \phi/\sqrt{7}-2 \hat \chi}}{\pi ^4  \beta v^{5/2} }\, , \\
 \frac{\partial \mathcal E}{\partial e_3} &= 0 \quad \rightarrow \quad e_3 = \frac{3  Q_3 e^{-2 \hat \phi/\sqrt{7}+2 \hat \chi}}{\pi ^4 \beta v^{5/2}}\, , \\
   \frac{\partial \mathcal E}{\partial v} &= \frac{\partial \mathcal E}{\partial \beta}= 0 \quad \rightarrow \quad v^6= \frac{3}{28 \pi ^{8}} \left( Q_1^2 e^{\frac{5 \hat \phi}{\sqrt{7}}-\hat \chi}+Q_2^2 e^{-\frac{2 \hat \phi}{\sqrt{7}}-2 \hat \chi}+Q_3^2 e^{-\frac{2 \hat \phi}{\sqrt{7}}+2 \hat \chi}\right), \ \beta = 36 \, ,\\
   \frac{\partial \mathcal E}{\partial \hat \chi} &= \frac{\partial \mathcal E}{\partial \hat \phi} =0 \quad \rightarrow \quad
 \hat \phi  = \frac{1}{4 \sqrt{7}}\log \left(\frac{2 Q_2^6 Q_3^2}{3 Q_1^8}\right) ,\quad \hat \chi = \frac{1}{4} \log \left(\frac{3 Q_2^2}{2 Q_3^2}\right)\,\label{eq:moduli} .
\end{align}
Finally, plugging all of the above into the entropy function we get the black hole entropy at the horizon
\begin{equation} \label{eq:entropy_9d}
{\mathcal E}|_h= 
S \sim \left| Q_1^2 Q_2^2 Q_3^3  \right|^{1/6} \, ,
\end{equation}
where we have ignored order one factors.

\paragraph{A black hole solution} \mbox{} \\
Given an extremum of the entropy function, the existence of a full-fledged black hole attractor solution corresponding to it is not always guaranteed. In \cite{Goldstein:2005hq}, it is proven that such a solution exists if the Hessian of the effective potential acting on the scalars has positive eigenvalues. This effective potential is related to the entropy function once it is extremized with respect to the geometry and only the scalar fields are left. The effective potential is written as:
\begin{eqnarray} \label{eq:truncated-effective-potential}
    V_{eff} \sim   Q_1^2 e^{-\hat{\chi}+5\hat{\phi}/\sqrt{7}} +Q_2^2 e^{-2\hat{\chi}- 2\hat{\phi}/\sqrt{7} }+Q_3^2 e^{2\hat{\chi}- 2\hat{\phi}/\sqrt{7} }\,.
\end{eqnarray}
We see that $\partial _ i V_{eff}=0$ at the horizon as dictated by \eqref{eq:moduli}, and one can check that the matrix $\partial_i \partial_j V_{eff}$ has positive eigenvalues for any (non-vanishing) charges.

Finding the most general black hole solution whose near-horizon geometry is described above is a daunting task. A simpler one to obtain is the so-called ``double-extreme'' solution where the moduli are constant along $r$ and are given by their attracted value. This solution is as follows: 
\begin{align}
    ds^2&= -\left( 1- \frac{r^6_H}{r^6}\right)^{2} dt^2 + \left( 1- \frac{r^6_H}{r^6}\right)^{-2} dr^2+ r^2 d\Omega_ 7^2\,, \\
    \hat \phi  &= \frac{1}{4 \sqrt{7}}\log \left(\frac{2 Q_2^6 Q_3^2}{3 Q_1^8}\right)   ,\label{eq:phisol} \\
	\hat \chi &= \frac{1}{4} \log \left(\frac{3 Q_2^2}{2 Q_3^2}\right) \,\label{eq:chisol} ,\\
 F_7^ {(i)} &= \star_{9d} F_2^{(i)} = Q_{(i)} d\text{vol}(S^7)\,,
\end{align}
with horizon
\begin{equation}
   r_H = \left( \frac{Q_1^{2} Q_2^{2} Q_3^{3}}{2^{16} 3^{5}} \right)^{\frac{1}{6\times 7} }\,.
\end{equation}
More general solutions, where the moduli have a non-trivial profile along $r$ and only reach their attracted value at the horizon can be studied by perturbing the solution above \cite{Goldstein:2005hq}.

\subsection{Minimal 9d Black Holes as UV Probes in Asymptotic Limits}
\label{sec:9d-limits-BH}

We now wish to study the behavior of the entropy \eqref{eq:entropy_9d} as a function of the moduli space distance along the various asymptotic limits. To do this, we examine the relation between the attracted value of the moduli and the charges given by \eqref{eq:phisol} and \eqref{eq:chisol}. In order to connect with a putative UV completion of the theory, we implicitly use the embedding in 9d supergravity i.e. of type II string theory on $\mathbb{S}^1$ or M-theory on $\mathbb{T}^2$.

\subsubsection{Decompactification to 11d Limit}
\label{sec:9d-limits-BH-decomp}

We first study the limit that corresponds to a decompactification to 11-dimensional M-theory. As we will check later, this can be achieved by taking
\begin{equation}
    Q_3 = \text{const.} \, , \quad Q_1 \sim Q_2 \sim Q \to \infty \, .
\end{equation}
The intuition behind this comes from blowing up the charges related to the leading tower in the limit we want to explore. Indeed, $Q_1$ and $Q_2$ are the charges of the KK modes to 11d M-theory.

Using \eqref{eq:moduli}, in this limit we get
\begin{equation} 
	\hat \phi \simeq - \frac{1}{2 \sqrt{7}} \, \log Q \, , \quad \hat\chi \simeq \frac{1}{2} \log Q \, .
\end{equation}
Taking into account that these are canonically normalized fields, this leads to the moduli space distance
\begin{equation}
    D_\phi \simeq \sqrt{\frac{2}{7}} \log Q \, .
\end{equation}
Using this result, together with \eqref{eq:entropy_9d} for the entropy, we then get that the black hole scale goes as
\begin{equation}
    \Lambda_{\rm BH} \sim S^{-1/7} \sim Q^{-2/21} \sim \exp \left( - \frac{1}{3} \sqrt{\frac{2}{7}} \, D_\phi \right) \, .
\end{equation}
It is a pleasant surprise that this fits the expectation for the species scale of a decompactification from $d=9$ to $D=11$ dimensions, i.e.
\begin{equation}
    \sqrt{\frac{D-d}{(d-2)(D-2)}} = \frac{1}{3} \sqrt{\frac{2}{7}} \, .
\end{equation}

Let us now check that the limit explored is indeed the decompactification to 11d M-theory. From \eqref{charge-states} below, and plugging the previous relations for the moduli, we obtain
\begin{equation}
    m_1 \sim m_2 \sim Q^{-3//7} \sim \exp \left( - \frac{3}{\sqrt{14}} \, D_\phi \right) \, .
\end{equation}
Indeed, we get two KK towers falling at the same rate, which coincides with the expected exponential rate for a decompactification from $d=9$ to $D=11$. Given the density of the two towers together, this yields the species scale \cite{Castellano:2021mmx}
\begin{equation}
    \Lambda_s \sim m^{2/9} \sim \exp \left( - \frac{1}{3} \sqrt{\frac{2}{7}} \, D_\phi \right) \, .
\end{equation}
We thus see explicitly that BH scale and the species scale match asymptotically in this decompactification limit to 11d M-theory. This a remarkable check that the results in our paper may persist beyond the realm of supersymmetric black holes, and that this pattern is not trivially explained by the added structure that supersymmetry imposes on the solution.

\subsubsection{An Emergent String Limit}
\label{sec:9d-limits-BH-string}

We now turn to the limit $\hat \phi \to \infty$ with $\hat{\chi}$ fixed. This corresponds to an emergent string limit, where the oscillation modes of the 10D string become light. In this picture (see e.g. \cite{Das:1996je}) the field $\chi$ parametrizes the radius square of the circle in string units, while $\hat{\phi}$ is the 9d dilaton. Taking this into account we get the following relations between scales:
\begin{equation} \label{eq:scales}
	 \frac{M_s}{M_{Pl,9}} \sim e^{\frac27 \phi} \sim e^{ -\frac{1}{\sqrt{7}} \, \hat \phi } \, , \quad \frac{M_{KK}}{M_s} \sim \chi^{-1/2} \sim e^{-\hat \chi}  \, .
\end{equation}
Hence, the limit $\hat \phi \to \infty$, $\chi$ fixed, corresponds to an emergent string limit. Indeed, one can recover the expected exponential rate for an emergent string limit \cite{Etheredge:2023odp} 
\begin{equation}
  \alpha = \frac{1}{\sqrt{d-2}} = \frac{1}{\sqrt{7}} \, .
\end{equation}

We now explore this asymptotic limit using the classical black holes above, in particular taking the following large charge limit $Q_2 \sim Q_3 \to \infty$. Using \eqref{eq:moduli}, this indeed fixes $\chi$ and explores
\begin{equation}
	 \hat \phi \simeq \frac{2}{\sqrt{7}} \, \log \left(  \frac{Q_2}{Q_1}  \right)  \to \infty\, .
\end{equation}
In this limit, the entropy scales as \eqref{eq:entropy_9d}
\begin{equation}
  S \sim \left| Q_1 \right|^{1/3} \left| Q_2 \right|^{5/6} \, .
\end{equation}
The smallest black holes in this family are obtained for the smallest $Q_1$, which leads to
\begin{equation}
  \Lambda_{\rm BH} \sim S^{-1/7} \sim \left| Q_2 \right|^{-5/42} \sim \exp \left( - \frac{5}{12\sqrt{7}}  \hat \phi \right) \, .
\end{equation}
Hence, this family of black holes does not reproduce the right exponential behavior given by the string scale in \eqref{eq:scales}. In fact, notice that these BHs become much smaller than the species length asymptotically. In Section \ref{sec:global picture}, we will study the exponential decay rate of $\Lambda_{\rm BH}$ along any infinite distance limit, precisely characterizing for which of them this mismatch with the species scale happens.

This fact that these black holes are parametrically smaller than the species length is surprising. It is in sharp contrast with the behaviour found in 4d and 5d top-down theories in sections \ref{sec:5d} and \ref{sec:4d}, where classical black holes are always larger than the species length, with at most minimal black holes saturating this bound. The resolution of this puzzle will be presented in section \ref{sec:puzzle-resolution}, and it lies in the inconsistency of the {\em ad hoc} removal of the axion, which is clearly present in the actual full 9d supergravity completion of the theory, and also plays a key role in the actual full solution of our black holes in supergravity. Before that, we continue the systematic exploration of black holes and their UV scales in our simple bottom-up model, taking advantage of its simplicity, to develop a global convex hull picture for the black hole UV scale.

\subsection{A Global Picture for Asymptotic Limits}
\label{sec:global picture}

We now briefly generalize our results for these 9d black holes for arbitrary asymptotic limits in moduli space. Being that there are only two moduli to consider, we can neatly package our results into a {\em black hole convex hull diagram}. This nicely puts the black hole UV scales in the arena of convex hull formulations of swampland conjectures \cite{Cheung:2014vva,Montero:2015ofa,Gendler:2020dfp,Calderon-Infante:2020dhm,Etheredge:2022opl,Calderon-Infante:2022nxb,Calderon-Infante:2023ler,Calderon-Infante:2024oed}.

\subsubsection{The Distance Conjecture and Species Scale Convex Hulls}
\label{sec:global-distance-conjecture-hull}

\begin{figure}[htb]
\begin{center}
\includegraphics[width= 0.5\textwidth]{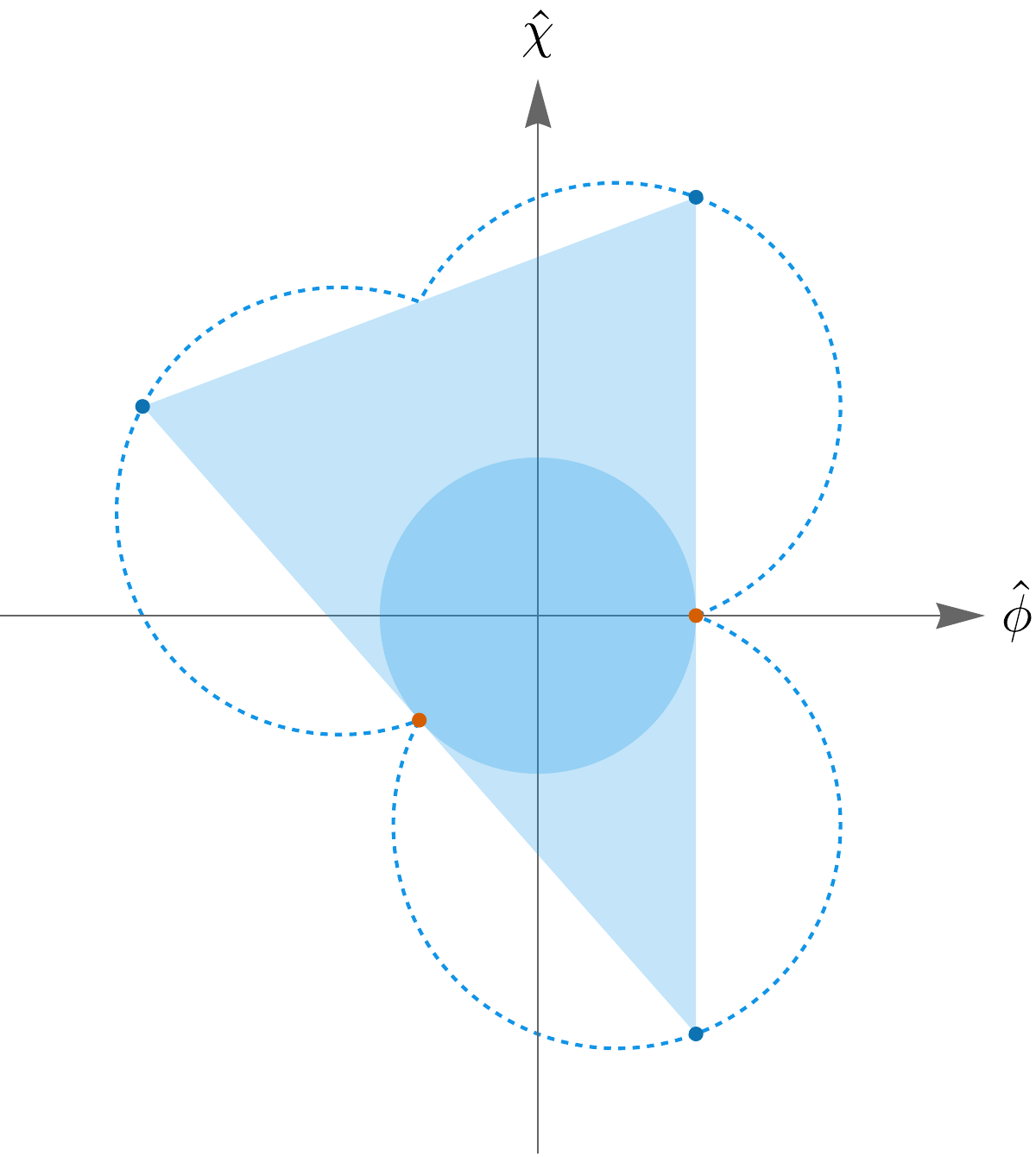}
\caption{\small Diagram for the distance conjecture convex hull and the asymptotic towers in the 9d maximal supergravity theory. The bubbly curve represents the exponential decay rate of towers (\ref{bubbly-rate}) in different directions in moduli space. The light blue triangle is the convex hull of the vectors \eqref{tower-vectors}, represented with blue dots. The orange dots describe the scalar charge to mass ratios for the string towers \eqref{string-vectors}. The darker blue circle represents the  sharpened Distance Conjecture bound, which is satisfied as it fits inside the convex hull triangle.} 
\label{fig:convex-hull-towers}
\end{center}
\end{figure}

Let us start by focusing on the scale of the different towers in the putative UV completion as 9d supegravity (i.e. type II on $\mathbb{S}^1$ / M-theory on $\mathbb{T}^2$). Reading off the gauge couplings from the kinetic terms, we see that the towers of BPS particles will behave as:
\begin{equation}
\begin{split} \label{charge-states}
	m_1 &\sim g_1 \sim s^{5/7} \chi^{-1/4} \sim \exp \left( \frac{5}{2 \sqrt{7}} \hat \phi - \frac{1}{2} \hat\chi \right) \, , \\
	m_2 &\sim g_3 \sim s^{-2/7} \chi^{-1/2} \sim \exp \left( - \frac{1}{\sqrt{7}} \hat \phi - \hat\chi \right) \, , \\
	m_3 &\sim g_3 \sim s^{-2/7} \chi^{1/2} \sim \exp \left( - \frac{1}{\sqrt{7}} \hat \phi + \hat\chi \right) \, .
\end{split}
\end{equation}
For our purposes, it will be useful to define the so-called scalar charge to mass ratios \cite{Calderon-Infante:2020dhm}
\begin{equation}
    \vec \zeta = - \vec \nabla \log M_{\rm t} \, ,
\end{equation}
where we are assuming a (locally) flat basis in moduli space. Since $\hat \phi$ and $\hat \chi$ are indeed flat coordinates, we can easily read off these vectors from \eqref{charge-states}:
\begin{equation} \label{tower-vectors}
	\vec \zeta_1 = \left( -\frac{5}{2 \sqrt{7}} \, , \, \frac{1}{2} \right) \, , \quad \vec \zeta_2 = \left( \frac{1}{\sqrt{7}} \, , \, 1 \right) \, , \quad \vec\zeta_3 = \left( \frac{1}{\sqrt{7}} \, , \, -1 \right) \, .
\end{equation}
Given the scalar charge to mass ratio of a tower, its exponential decay rate along a trajectory with unit tangent vector $\hat T$ can be computed as
\begin{equation}
    \alpha_{\rm t}(\hat T) = \hat T \cdot \vec \zeta_{\rm t} \ .
\end{equation}
When the scalar charge to mass ratios have non-trivial dependence in moduli space, we are implicitly taking them to the regime dictated by the tangent vector $\hat T$. Finally, the asymptotically lightest tower along a given direction will correspond to the larger $\alpha_{\rm t}(\hat T)$. Its exponential decay rate is then given by
\begin{equation}
    \alpha(\hat T) = \max \left\{ \alpha_{\rm t}(\hat T) \right\} \, .
    \label{bubbly-rate}
\end{equation}
Performing a polar plot of this quantity for the vectors in \eqref{tower-vectors}, we get the blue bubbly curves in figure \ref{fig:convex-hull-towers}, where the vectors correspond to the blue dots. In addition, we also display their convex hull (light blue triangle) together with the sharpened bound for the Distance Conjecture \cite{Etheredge:2022opl} (dark blue circle). For completeness, we also highlight with red dots the scalar charge to mass ratios for the string towers
\begin{equation} \label{string-vectors}
    \vec \zeta_{s1} = \left( \frac{1}{\sqrt{7}} \, , \, 0 \right) \, , \quad \vec \zeta_{s2} = \left( - \frac{3}{4\sqrt{7}} \, , \, - \frac{1}{4} \right) \, .
\end{equation}
As proposed in \cite{Calderon-Infante:2020dhm}, the fact that the convex hull contains the ball representing the sharpened bound guarantees that the latter is satisfied for any asymptotic limit. Incidentally, we recover a rotated version of the convex hull in \cite{Etheredge:2022opl} (figure 4 therein). This is a non-trivial check of the gauge kinetic functions above.

\begin{figure}[htb]
\begin{center}
\includegraphics[width= 0.5\textwidth]{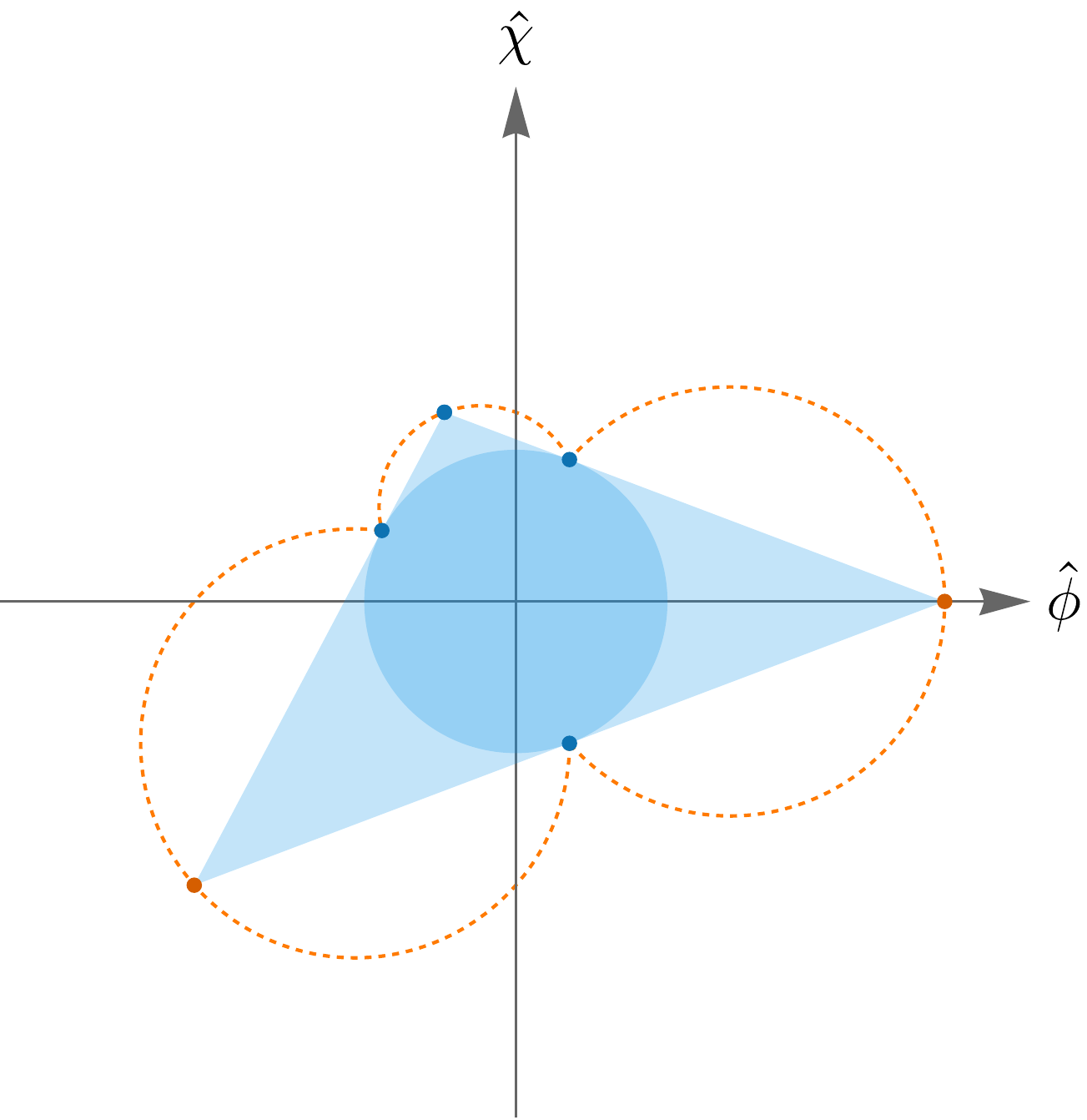}
\caption{\small Diagram for the species scale convex hull in the 9d maximal supergravity theory. The bubbly curve represents the exponential decay rate of the species scale in different directions in moduli space. The light blue triangle is the convex hull of the species vectors \eqref{species-vectors}, represented with blue and orange dots, for KK and string towers, respectively. The darker blue circle represents the  Species Scale Distance Conjecture bound, which is satisfied as it fits inside the convex hull triangle.} 
\label{fig:convex-hull-species}
\end{center}
\end{figure}

A similar figure can be obtained for the exponential decay rate of the species scale. As explained in \cite{Calderon-Infante:2023ler}, the vectors in \eqref{tower-vectors} and \eqref{string-vectors} ---together with the information that they correspond to KK and string towers, respectively--- leads to the following \emph{species vectors}:
\begin{equation} \label{species-vectors}
\begin{split}
    \vec {\mathcal Z}_{1} = \left(-\frac{5}{16 \sqrt{7}},\frac{1}{16}\right) \, , \quad \vec {\mathcal Z}_{2} = \left(\frac{1}{8 \sqrt{7}},\frac{1}{8}\right) \, , \quad  \vec {\mathcal Z}_{1} = \left(\frac{1}{8 \sqrt{7}},-\frac{1}{8}\right) \, , \\
    \vec {\mathcal Z}_{(12)} = \left(-\frac{1}{6 \sqrt{7}},\frac{1}{6}\right) \, , \quad \vec {\mathcal Z}_{s1} = \left( \frac{1}{\sqrt{7}} \, , \, 0 \right) \, , \quad  \vec {\mathcal Z}_{s2} = \left( - \frac{3}{4\sqrt{7}} \, , \, - \frac{1}{4} \right) \, .    
\end{split}
\end{equation}
These vectors encode the asymptotic exponential decay rate of the species scale in the same way that the $\vec \zeta$-vectors encode that of the towers. The information for the species scale along different directions can be displayed in a plot analogous plot to Figure \ref{fig:convex-hull-towers}, which we provide in Figure \ref{fig:convex-hull-species}. In this case, the (dark blue) ball represents the Species Scale Distance Conjecture bound \cite{Calderon-Infante:2023ler} (see also \cite{vandeHeisteeg:2023uxj}). As a check, we directly see that we recover a rotated version of the convex hull in \cite{Calderon-Infante:2023ler} (figure 8 therein).

\subsubsection{The Black Hole Convex Hull}
\label{sec:global-black-hole-hull}

In this section, we obtain a similar figure for the scale explored by minimal BHs in each asymptotic limit in the 9d EFT obtained by truncating away the axion. In Section \ref{sec:9d-limits-BH}, we first chose the asymptotic limit by fixing some ratio of charges and then restricted to the minimal BH. For our purposes in this section, it will be more convenient to first restrict to a set of minimal BHs and then consider all the asymptotic limits explored by them. 

It is apparent from \eqref{eq:entropy_9d} that the BH entropy ---and therefore the size--- grows with the three charges $Q_1$, $Q_2$ and $Q_3$. On the other hand, we see from \eqref{eq:moduli} that the BH horizon will explore an asymptotic limit only if one or more of the charges blow up. In particular, since we want to be able to change the rate at which $\hat \phi$ and $\hat \chi$ grow or decrease, we need to be able to play with at least two of the charges. Following this reasoning, we conclude that the way of choosing a family of \emph{minimal} BHs while keeping the freedom of tuning the asymptotic limit is to fix one of the charges while taking the other two to infinity at different rates.

\begin{figure}[htb]
\begin{center}
\includegraphics[width= 0.5\textwidth]{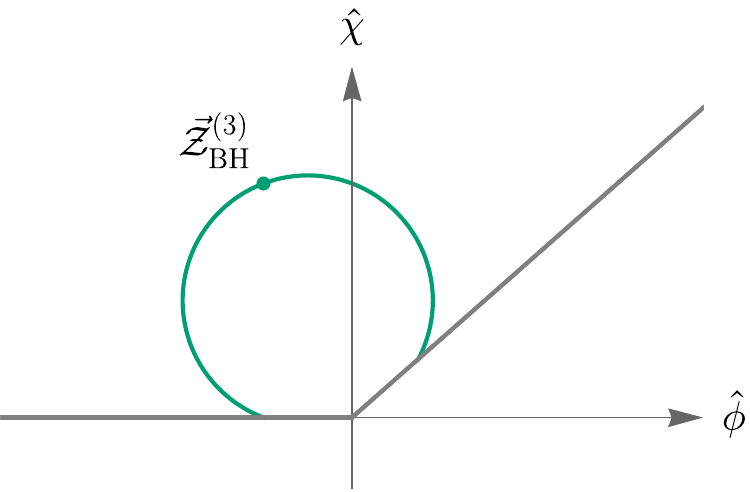}
\caption{\small Polar plot of $\hat T \cdot \vec {\mathcal Z}_{\rm BH}^{(3)}$ (green curved line). We restrict to the range of $\hat T$ explored by black holes with $Q_3 \sim \mathcal O(1)$, i.e., those leading to $\vec {\mathcal Z}_{\rm BH}^{(3)}$.} 
\label{fig:BH-bubbly-3}
\end{center}
\end{figure}

Let us first fix $Q_3 \sim \mathcal{O}(1)$. Plugging into \eqref{eq:moduli} and inverting the relations, we obtain
\begin{equation}
    |Q_1| \sim e^{-\frac{\sqrt{7}}{2} \hat\phi + \frac{3}{2} \hat\chi} \, , \quad |Q_2| \sim e^{2 \hat\chi} \, .
\end{equation}
Using this expression and \eqref{eq:entropy_9d}, the BH scale then reads
\begin{equation} \label{eq:BH-vector}
    \Lambda_{\rm BH} \sim S_{\rm BH}^{-1/7} \sim e^{\frac{1}{6\sqrt{7}} \hat\phi - \frac{1}{6} \hat\chi } \, ,
\end{equation}
from which we easily read off the \emph{BH vector}
\begin{equation}
    \vec {\mathcal Z}_{\rm BH}^{(3)} \equiv - \vec\nabla \log  \Lambda_{\rm BH} = \left( -\frac{1}{6\sqrt{7}} , \frac{1}{6} \right) \, .
\end{equation}
Incidentally, we recover the species vector $\vec{\mathcal Z}_{(12)}$ in \eqref{species-vectors}. Let us however point out that this vector is not the right one to compute the exponential decay rate of the BH scale for any asymptotic limit. To see this, let us take
\begin{equation}
    Q_1 \sim Q^{p_1} \, , \quad Q_2 \sim Q^{p_2} \, , \quad Q\to \infty \, ,
\end{equation}
where $Q,p_1,p_2 \in \mathbb{N}$ so that charge quantization is satisfied. Plugging this into \eqref{eq:moduli}, we obtain
\begin{equation}
    \hat \phi = \frac{-4 p_1 + 3 p_2}{2 \sqrt{7}} \log Q \, \quad \hat \chi = \frac{p_2}{2} \log Q \, .
\end{equation}
This allows us to explore different asymptotic limits as expected. The unit tangent vector of the trajectory explored by this family of BHs is given by
\begin{equation}
    \hat T = \left( \frac{3-4 r}{2 \sqrt{4 r^2-6 r+4}},\frac{\sqrt{7}}{2 \sqrt{4 r^2-6 r+4}}\right) \, ,
\end{equation}
where we have defined the ratio $r=p_1/p_2$. Notice that, even though $p_1,p_2 \in \mathbb{N}$, the ratio $r$ densely populates the range $r\in [0,\infty)$ and thus can be regarded as a continuous variable. For this range of $r$, the angle of the tangent vector to the $\hat \phi$ axis is bounded between $\arctan(\sqrt{7}/3)$ and $\pi$. Hence, for any trajectory with unit tangent vector within this range, we can use \eqref{eq:BH-vector} to compute the exponential decay rate of the BH scale as $\hat T \cdot \vec {\mathcal Z}_{\rm BH}^{(3)}$. Performing a polar plot of this quantity ---restricted to its regime of validity--- we obtain Figure \ref{fig:BH-bubbly-3}.

Performing the same exercise for $Q_1 \sim \mathcal{O}(1)$ and $Q_2 \sim \mathcal{O}(1)$, respectively, we obtain the following other two BH vectors:
\begin{equation} \label{BH-vectors-2}
    \vec {\mathcal Z}_{\rm BH}^{(1)} = \left( \frac{5}{12 \sqrt{7}},-\frac{1}{12}\right) \, , \quad \vec {\mathcal Z}_{\rm BH}^{(2)} = \left( -\frac{1}{6 \sqrt{7}},-\frac{1}{6}\right) \, .
\end{equation}
By a similar reasoning as the one above, these two vectors turn out to be valid in the ranges $\theta \in (-\arctan(\sqrt{7}),\arctan(\sqrt{7}/3))$ and $\theta \in (-\pi,-\arctan(\sqrt{7}))$, respectively. Notice that, together with the range of validity of the $Q_3 \sim \mathcal{O}(1)$ case, this completes the full range of possible tangent vectors. Thus, by using the right BH vector, we can now compute the exponential decay rate of the BH scale as $\hat T \cdot \vec {\mathcal Z}_{\rm BH}^{(i)}$. 

\begin{figure}[htb]
\begin{center}
\includegraphics[scale=.5]{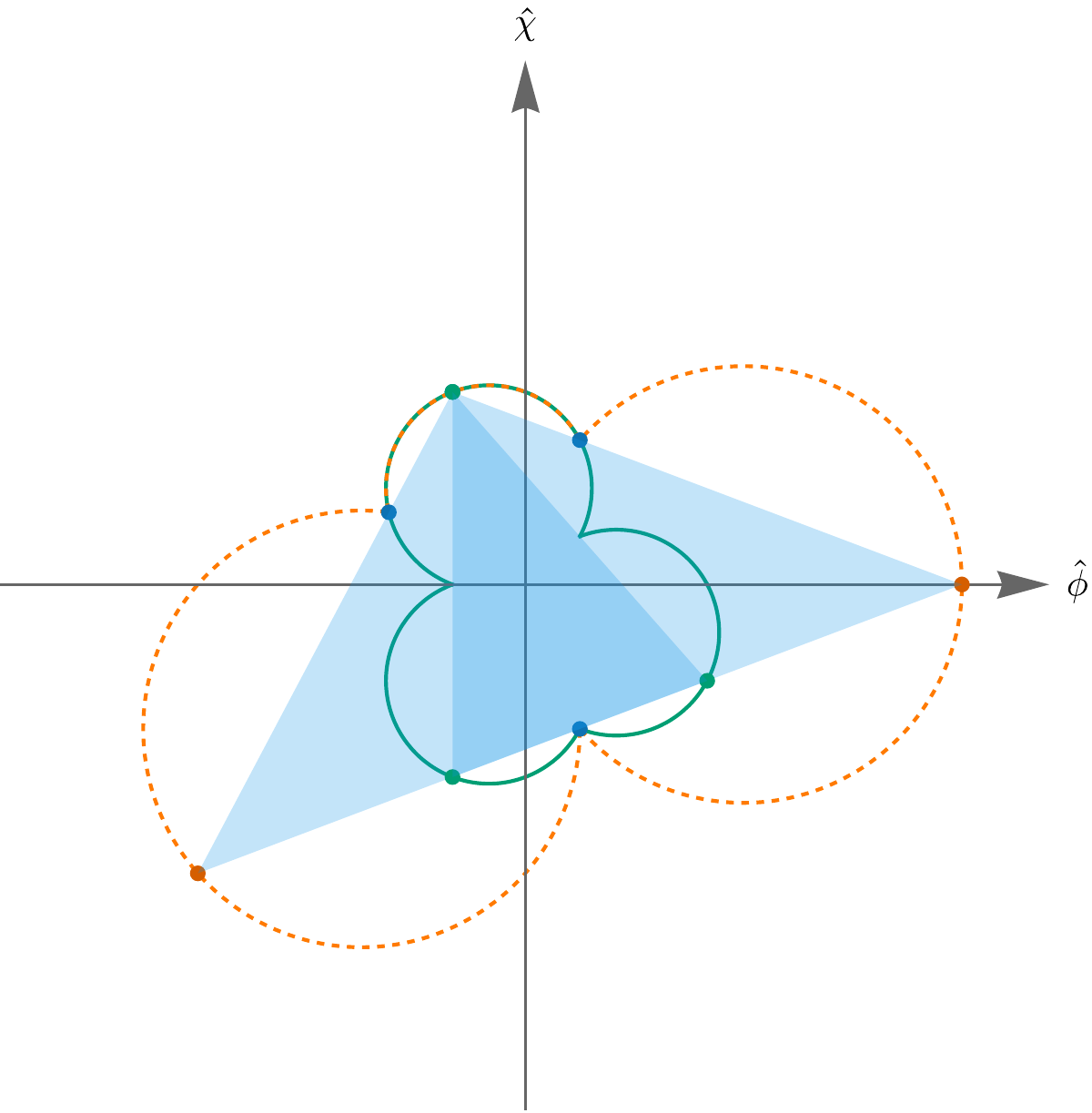}
\caption{\small Comparison of the exponential decay rates of the black holes scale (green) and the species scale (dashed and orange). For completeness, the convex hulls of the BH and species vectors are also shown. } 
\label{fig:convex-hull-9d}
\end{center}
\end{figure}

Performing a polar plot as the one in Figure \ref{fig:BH-bubbly-3} for the BH vectors in \eqref{BH-vectors-2} ---restricted to their regime of validity above--- and gluing together the resulting diagrams together with Figure \ref{fig:BH-bubbly-3}, we obtain the green line in Figure \ref{fig:convex-hull-9d}. The convex hull of the BH vectors is also shown for completeness. For comparison, we also show the  exponential decay rate and the convex hull for the species scale (c.f. Figure \ref{fig:convex-hull-species}). \\

Quite remarkably, we see that the BH scale generically deviates from the species scale, with the only exception of pure KK decompactification limits like the one considered in section \ref{sec:9d-limits-BH-decomp}. Along essentially any other direction, the minimal black holes turn out to be parametrically smaller than the species length in the putative UV completion of the theory. This is in stark contrast with the behavior in 5d and 4d theories, where classical black holes are always larger than this scale, with at most minimal ones saturating this bound. In the next section we explain that this pathology is indeed a reflection of the incompleteness of the truncated 9d EFT theory and hence of the black hole solutions as described above.

\subsection{Resolution of the Puzzle}
\label{sec:puzzle-resolution}
 
In the previous section, we analyzed the behavior of black holes constructed in the bottom-up 9d EFT as they probed the boundary of moduli space. We found that these minimal black holes only follow the species scale in decompactification limits, whilst they completely underestimate the species length in the other limits. In this section, we now clarify this apparent puzzle.

Recall that our bottom-up 9d theory was obtained as a  truncation of 9d maximal supergravity, by setting the axion $a$ to zero in \eqref{eq:fullact}. In this section, we put the axion back in to complete the bottom-up 9d EFT action to maximal supergravity, which is the theory that actually allows for an UV completion into type II on $\mathbb{S}^1$ / M-theory on $\mathbb{T}^2$. We show that the black hole solutions of the truncated theory do not survive as classical finite-sized black holes upon including the axion.

Allowing for a non-vanishing axion value, the effective potential in \eqref{eq:truncated-effective-potential} becomes
\begin{eqnarray} \label{eq:effective-potential}
    V_{eff} \sim Q_1^2 e^{-\hat{\chi}+5\hat{\phi}/\sqrt{7}} +Q_2^2 e^{-2\hat{\chi}- 2\hat{\phi}/\sqrt{7} }+(Q_3 - a Q_1)^2 e^{2\hat{\chi}- 2\hat{\phi}/\sqrt{7} }\,.
\end{eqnarray}
The derivative with respect to the axion yields
\begin{equation}
    \partial_a V_{eff} = 2 e^{2\hat{\chi}- 2\hat{\phi}/\sqrt{7}} Q_1 (a Q_1 - Q_3) \, . 
\end{equation}
Hence, extremization of the effective potential with respect to the axion requires
\begin{equation}
    a = \frac{Q_3}{Q_1} \, .
\end{equation}
This shows that setting the axion to zero as in the truncated theory is not consistent unless $Q_3=0$. Furthermore, plugging this condition back into the effective potential in \eqref{eq:effective-potential}, we see that the last term vanishes and the effective potential becomes
\begin{equation}
    V_{eff} \sim Q_1^2 e^{-\hat{\chi}+5\hat{\phi}/\sqrt{7}} + Q_2^2 e^{-2\hat{\chi}- 2\hat{\phi}/\sqrt{7} } \, . 
\end{equation}
Taking derivatives with respect to the saxions, one finds no valid extremum for this effective potential. All in all, including the axion field leads to an extra condition that renders the system of equations for extremizing the effective potential incompatible. Thus, the full 9d maximal supergravity theory does not allow for finite-sized classical black holes. 

As advertised, the black holes of the truncated 9d theory ---including those parametrically smaller than the species length--- do not survive as finite-sized black holes after including the axion. We take this as evidence that the bottom-up 9d EFT cannot be consistently UV-completed due to the lack of the axion field, which is indeed crucial to embed it into type II on $\mathbb{S}^1$ / M-theory on $\mathbb{T}^2$. This gives further support to the idea that classical minimal black holes size are bounded by the species length in theories that can be UV-completed to quantum gravity. It would be interesting to further our understanding of this connection, and perhaps even use this idea as a Swampland criterion to distinguish theories that can be UV-completed to quantum gravity from those that cannot. We leave this as an interesting direction for future work, and now turn towards possible explanations for the fact that classical black holes know about UV scales.

\section{Towards a Microscopic Interpretation}
\label{sec:interpretation}

We have shown that minimal size classical black holes, built using the 2-derivative action, can encode data about UV scales. This is a remarkable result, that hints at some notion of UV/IR mixing and to the broad idea that the 2-derivative low-energy EFT knows something about the UV. In this section, we explore possible arguments to explain this behaviour. First, in section \ref{sec:tracking}, we will explain what constrains the existence of such classical black hole solutions. In particular, we will argue that the UV data enters through the specific structure of the gauge kinetic functions. In section \ref{sec:micro} we will explore explanations in terms of microstate counting and its interplay with the states in the asymptotic towers.

\subsection{Tracking Down the UV Data}
\label{sec:tracking}

We now explore the reason why the classical black hole families described in the previous sections sometimes seem to follow UV scales from considerations of the effective action and its origin via emergence. We consider the Einstein-Maxwell-Dilaton sector of a general $d$-dimensional theory with $n$ $U(1)$ gauge fields, and $m$ real scalars. The gauge fields couple to the scalars through a gauge-kinetic matrix, which we will assume to be diagonal for simplicity. We write the action as: 
\begin{equation}
    S= \frac{1}{2\kappa^2}\int d^d x \sqrt{-G}\{R - g_{ij} \partial _ \mu \phi^i \partial ^ \mu \phi^ j - \frac{1}{4} \sum_{i=1}^n e^{f_i(\phi)}(F^{(i)}_2)^2 \} \, .
\end{equation}
We assume the existence of an extremal charged black hole solution, with near-horizon ansatz
\begin{equation}
\begin{split} 
      ds^{2} &= \frac{v}{\beta} \left( -r^2 dt^2 + \frac{1}{r^2} dr^2 \right) + v \, d\Omega_{d-2}^{2} \, , \\
	  F_{rt}^{(i)} &= e_i \, , \partial_\mu \phi^i = 0 \,  .\\
\end{split}
\end{equation}
This black hole background must be a solution to the extremization of the entropy function \cite{Sen:2005wa}
\begin{equation} 
  \mathcal E (Q_i,e_i,\phi^i,v,\beta) = 2\pi \left(  \sum_{i=1}^{n} e_i Q_i - \int_{S^{d-2}} d\Omega_{d-2} \sqrt{-G} \, \left.\mathcal L\right|_{h} \right) \, .
\end{equation}
Evaluating this entropy function at the horizon is straightforward, using: 
\begin{align}
    R|_h&= \frac{-2 \beta + (d-3)(d-2)}{v} &\sqrt{-G}|_h&= \frac{vol\; v^{d/2}\, d}{\beta}\\
    (F_2^{(i)})^2|_h &= - \frac{2 \beta^2 e_i^2}{v^2}& \partial \phi^i |h&=0 
\end{align}
where $vol$ is the volume of the unit $(d-2)$-sphere. As we did in Section \ref{sec:class-9DBHs}, we can evaluate the entropy function at the horizon and perform the extremization with respect to $e_i$ and the geometric parameters $\beta$ and $v$:
\begin{align}
     \frac{\partial \mathcal E}{\partial e_i} &= 0 \quad \rightarrow \quad e_i = \frac{e^{- f_i(\phi)}Q_i}{v^{\frac{d-4}{2}}  \beta\; vol}\, ,\label{eq:genmine} \\
   \frac{\partial \mathcal E}{\partial v} &= \frac{\partial \mathcal E}{\partial \beta}= 0 \quad \rightarrow \quad v^{d-3}= \frac{\sum_{i=1}^n Q_i^2 e^{-f_i(\phi)}}{2(d-3) (d-2) \;vol^2}\, , \ \beta = (d-3)^2 \, .\label{eq:genminnub}
\end{align}

So far, we have made no assumptions about the dependence of the gauge couplings on the moduli. Actually, it is crucial that the functions $f_i(\phi)$ are linear in the canonically normalized moduli for the entropy of such classical black holes to go exponentially with them. Indeed, one can plug equations \eqref{eq:genmine} and \eqref{eq:genminnub} into the entropy function and obtain:
\begin{equation}
    S\sim \mathcal{E}|_h \sim \left(\sum_{i=1}^n e^{-f_i(\phi)}Q_i^2\right)^{\frac{d-2}{2(d-3)}}\,.
\end{equation}
From this relation, we see that, if the functions $f_i(\phi)$ are linear in the canonically normalized moduli, then the entropy naturally follows a scaling $S \sim e^{\gamma \phi}$, with $\gamma$ some function that depends on the dimension and on the specific $\mathcal{O}(1)$ parameters in the functions $f_i(\phi)$. Indeed, consider writing $f_i(\phi)= \sum_{j=1}^m \alpha_{ij} \phi^j $, with $\alpha_{ij} $ some $\mathcal{O}(1)$ parameters. Then extremizing with respect to the moduli will lead to $m$ equations of the form:
\begin{equation}\label{eq:fullsol}
     \frac{\partial \mathcal E}{\partial \phi^j} = 0 \quad \rightarrow \quad  \sum_{i=1}^n \alpha_{ij} e^{-f_i(\phi)}Q_i^2 =0 \,.
\end{equation}
Rewriting the moduli as $\phi^j = \log x^j$, one can interpret this problem as finding the roots of $m$ polynomial equations in the $m$ variables $x^j$, supported on the same monomials, with coefficients given by the charges times some $\alpha_{ij} $. Determining when these equations have a solution is difficult, though not necessarily impossible (for instance using a version of Newton's identities to examine how the roots of a polynomial depend on its coefficients). However, the existence of a solution depends heavily on the precise parameters $\alpha_{ij}$, and the precise characterization of this is beyond the scope of this work. Happily, for our purposes it suffices to note that the solutions will generically lead to the moduli $\phi^j$ going as a logarithm of some homogeneous function of degree 0 in the charges. This in particular means that re-scaling all of the charges by a number does not change the attractor point in moduli space. It also implies that the charges which diverge as the attractor point is pushed to the boundary of moduli space will go as an exponential of the moduli. This in turn shows that the entropy will follow an exponential scaling law  $S \sim e^{\gamma \phi}$, where $\phi$ parametrizes some direction in the boundary of moduli space. 

Determining the exact value of $\gamma$ is the difficult part and highly depends on the specific black hole solution at hand, after solving the equations \eqref{eq:fullsol}. However, it is easy to see from these general considerations that, for a given theory and a given infinite distance limit parametrized by $\phi$, there will be a lower bound on the $\gamma$ given by the smallest such family of black holes. Whether or not this $\gamma$ follows a UV scale such as the species scale or the KK scale is a question of whether the $\alpha_{ij}$ conspire to yield the right dependence of the entropy on the moduli, after the full solution is obtained. This seems highly non-trivial, but we have shown that it actually does happen in large classes of BPS black holes. This remarkable fact begs for an explanation, which in the following we argue is related to emergence.

They key ingredient in the above argument is that the gauge gauge kinetic functions are exponentials in the canonically normalized moduli. The importance of this exponential dependence of gauge kinetic functions on moduli in quantum gravity theories has already been noted in the Swampland program, see e.g. \cite{Grimm:2018ohb,Heidenreich:2018kpg,Palti:2019pca,Gendler:2020dfp,vanBeest:2021lhn,Hamada:2021yxy,Marchesano:2022axe,Castellano:2022bvr,Angius:2023xtu,Blumenhagen:2023tev}. In particular, for such dilatonic couplings several formulations of proposed swampland conjectures, such as the tower Weak Gravity Conjecture \cite{Heidenreich:2015wga,Heidenreich:2016aqi,Andriolo:2018lvp}, the Repulsive Force Conjecture \cite{Heidenreich:2019zkl} and the Distance Conjecture, become equivalent \cite{Lee:2018spm,Gendler:2020dfp}. An even more relevant fact for our purposes is that this exponential dependence on moduli has been argued to arise via emergence from integrating out the states in the asymptotic tower \cite{Grimm:2018ohb,Heidenreich:2018kpg} (see also \cite{Palti:2019pca,vanBeest:2021lhn,Marchesano:2022axe,Castellano:2022bvr,Blumenhagen:2023tev} for related results). 

This hence suggests a plausible explanation for the fact that classical black hole solutions of the two-derivative EFT can encode information about UV scales and asymptotic towers. Namely, the latter produce via emergence the gauge kinetic function, and the resulting  exponential dependence of the latter on moduli precisely allows the families of minimal classical black holes to track UV scales via their entropy. 

We find this interlocked set of relations extremely satisfying. On the other hand, notice that the argument above works for charged black holes. It would be interesting to develop a similar understanding for neutral black holes, possibly along the lines of \cite{Bedroya:2024uva}. Leaving this question for future work, we now turn to a complementary discussion of the recovery of UV scales for charged black holes from the viewpoint of their microstate composition.

\subsection{Microstate Argument}
\label{sec:micro}

In this section, we explore arguments related to the  microstate structure of the black holes to explain their capability to probe UV scales of the theory despite their formulation in pure EFT terms. For concreteness, we focus on the case of 4d black holes, and center the discussion in the asymptotic regime, specifically of Type IV and II, in which classical black holes reproduce the species scale. The original definition of the species scale in asymptotic limits is in terms of the tower of species becoming light in the limit. Following our heuristic argument in section \ref{sec:4d}, the dominant composition (i.e. the diverging charges) of the black holes reproducing the species scale belong to these towers. Hence, we can explore the appearance of the species scale in terms of the microstate composition of these black holes. In particular, following the diverging charges reviewed in section \ref{sec:4d-limits-BH}, for Type IV asymptotic limits we deal with light species tower of D0-branes, while for Type II limits the family of black holes reproducing the species scale is that with D4/D0-brane charges.

Let us start with the Type IV limits. The asymptotic limit corresponds to a decompactification of the ${\mathbf S}^1$ in the description of M-theory on $\mathbf{X}_6\times {\mathbf S}^1$, so it makes sense to describe the family of black holes in M-theory terms. The black holes are microscopically described in terms of M5-branes wrapped on a 4-cycle, and carrying momentum in the ${\mathbf S}^1$. These black holes were considered in detail in \cite{Maldacena:1997de}, which provided a microstate counting reproducing their entropy, as follows.\footnote{In \cite{Maldacena:1997de} the microscopic match is reproduced even including the higher-curvature corrections. In the spirit of this work, we simply review the argument at the 2-derivative level.} 
The black hole D4-brane charge corresponds to an M5-brane wrapped on a 4-cycle  $P=p^IB_I$ in a basis of 4-cycles, times the ${\mathbf S}^1$, and carrying $q_0$ units of momentum. The reduction of the M5-brane on $P$ gives an effective string in the 5d theory, whose worldsheet theory is a $\mathcal N =(0,4)$ sigma model for the M5-brane moduli. The detailed description of the latter showed the 2d worldsheet left-moving central charge to be given by the self-intersection of the 4-cycle $P$, namely
\begin{equation}
c_L=C_{IJK}p^I p^J p^K \, .
\end{equation}
As in \cite{Strominger:1996sh}, the microscopic entropy of the system is given by the logarithm of the number of left-moving excitations with total momentum $q_0$, which for large $q_0\gg c_L$ is
\begin{equation}
S=2\pi \sqrt{\frac{c_L q_0}{6}} \, .
\label{msw-entropy}
\end{equation}

The above result is usually derived using Cardy's formula \cite{Cardy:1986ie}, but we are interested in a more
constructive explanation from combinatorics, as follows. The 2d theory has a number $c_L$ of degrees of freedom, which we can think of as free bosons, which can be expanded in terms of oscillators $\alpha_n^a$, $a=1,\ldots, c_L$. For $c_L=1$, the number of states in the sector of charge $q_0$ is given by the number of partition $p(q_0)$ of the momentum among the different oscillators of the momentum carried. For large $q_0$, the Hardy-Ramanujan asymptotic formula gives
\begin{equation}
p(n)\sim \frac{1}{4 n \sqrt{3}} \exp ( 2\pi \sqrt{n/6})
\end{equation}
in agreement with the entropy (\ref{msw-entropy}) for $c_L=1$. The asymptotic formula can be shown using the generating function for partitions
\begin{equation}
\sum_n p(n) q^n = \prod_{n=1}^{\infty}(1-q^n)\equiv  q^{\frac 1{24}}\eta^{-1}(\tau)\,
\end{equation}
and applying the Tauberian theorem of Ingham \cite{Ingham:1941}, see e.g. \cite{Rolen}. The latter relies on modular properties of the function, so it can be regarded as a mathematical avatar of Cardy's formula.

For general $c_L$, the argument is similar, but the combinatorial problem is the counting of $c_L$-colored partitions of $q_0$, namely partitioning among $c_L$ carriers. The generating function of the number $p^{(r)}(n)$ of $r$-colored partitions of $n$ is
\begin{equation}
\sum_n p^{(r)}(n) q^n = \prod_{n=1}^{\infty}(1-q^n)^{-r}\equiv \left(\, q^{\frac 1{24}}\eta^{-1}(\tau)\,\right)^r
\end{equation}
and using the above techniques, the asymptotic formula is
\begin{equation}
p^{(r)}(n) \sim p_r(n)\sim \exp ( 2\pi \sqrt{rn/6}) \, ,
\label{asymptotic-colored-partitions}
\end{equation}
which for $r=c_L$ and $n=q_0$ reproduces the entropy (\ref{msw-entropy}). The result (\ref{asymptotic-colored-partitions}) can also be obtained directly as a particular case of Theorem 1.1 in \cite{Chen-Li:2016}.

The above explanation of the black hole entropy in terms of a microscopic counting of partitions allows to describe the species scale as follows (see \cite{Calderon-Infante:2023uhz} for the argument applied to small black holes, i.e. of zero-size at classical level). The D4/D0 black hole family can be regarded as different dressing of a D4-brane state with different sets of D0-brane bound states, adding up to total charge $q_0$. We now recall the definition of the species length as the size of the smallest possible black hole in a given EFT \cite{vandeHeisteeg:2023ubh}. One might expect that, in order for the EFT to be able to describe the entropy of a charge $q_0$ black hole, it should include at least $q_0$ species (i.e. for the different bound states up to $q_0$ D0-branes). However, this is not correct and highly overestimates the minimum number of species actually necessary to explain the entropy, which we now show is given by $N_s=\sqrt{c_Lq_0}$. 

We start with the case $c_L=1$, so we deal with usual partitions of $q_0$. If we consider a theory with the number of species given by $k$ (i.e. bound states up to $k$ D0-branes) and want to build a charge $q_0$ black hole, the number of ways to do so is given by $p_k(q_0)$, the number of partitions of $q_0$ with each part {\rm not larger} than $k$. This is equivalent to the number of partitions of $q_0$ into at most $k$ parts, for which the classic result \cite{erdos-lehner} implies that $p(q_0)\sim p_k(q_0)$ in the limit $q_0\to \infty$ for $k\sim \sqrt{q_0}$. In other words, it suffices to have species up to  $N_{sp}=\sqrt{q_0}$ to dress the D4-brane state in a large enough number of ways to explain the full black hole entropy. In fact, the black hole entropy is dominated by states corresponding to partitions with one bound state of {\em exactly} $N_s=\sqrt{q_0}$ D0-branes, we refer the reader to \cite{Calderon-Infante:2023uhz} for details.

The above arguments extend easily to the general $c_L$ case. In this case, the entropy of the charge $q_0$ black hole is dominated by states in which $q_0$ is equally distributed among the $c_L$ carriers, and so we need to partition $q_0/c_l$ for each carrier. The distribution of these latter partitions is dominated by states with $\sqrt{q_0/c_L}$ D0-branes, so the total number of species, including the $c_L$-fold multiplicity is given by $N_{sp}=\sqrt{c_Lq_0}$, reproducing the black hole species scale.

The heuristic interpretation is that the black hole families with divergent charges are actually exploring the tower of light species arising in the corresponding asymptotic limit, and their corresponding classical entropies encode the relevant information of the underlying microscopic counting problem, hence are capable of reproducing the species scale from purely classical considerations.

A related comment is that, in the above combinatorial arguments we have essentially used the D0-brane charge, while the D4-brane charge plays more of an spectator role. In fact, the counting problems turn out to be essentially identical to those for zero-size black holes in \cite{Calderon-Infante:2023uhz}, so they lead to the same value of the species scale. Remarkably, while in \cite{Calderon-Infante:2023uhz} the actual species scale is reproduced by the stretched horizon obtained upon including higher-curvarture corrections, in the present case it is reproduced by the classical horizon which arises due to the presence of the additional finite D4-brane charge. In other words, although the family of minimal black holes probes the species scale, they are four-charge black holes (only one of the four charges diverges at infinity) with a macroscopic horizon at classical level.

Finally, we stress that the above combinatorial arguments only work in the case of a Type IV limit, where the number of D0 branes is much larger than that of D4 branes, and this limit is realized by the Cardy regime $q_0\gg c_L=C_{IJK}p^I p^J p^K$. In the Type II limit however, the number of D0 branes, $q_0$, scales the same way as that of the D4 branes wrapping the fiber, $p^1$. The black hole entropy \eqref{entropy_typeII_D0D4D4D4}, which scales like $\sim \sqrt{q_0p^1}$, is not dominated by states in which the D0-brane charge $q_0$ is equally distributed among the D4 branes or D4-brane self-intersections that act independently. Rather, one needs all the $q_0$ D0 branes and the $p^1$ D4 branes wrapping the fiber to interact together to produce $\sim e^{\sqrt{q_0p^1}}$ bound states. 

Since the number of states that make out this entropy scales like $\sim \sqrt{q_0p^1}$, it is tempting to think of them as D0-D4 bound states with D0- and D4-brane charges smaller than $(\sqrt{q_0},\sqrt{p^1})$. However, this cannot be correct, since the entropy \eqref{entropy_typeII_D0D4D4D4} comes from the number of partitions of $q_0p^1$ instead of the number of vector partitions of $(q_0,p^1)$. In other words, the number of such states depends only on the product $q_0p^1$, and not on the specific value $q_0$ and $p^1$ take. In the T-dual D1-D5 language, the microstates of the two-charge D1-D5 black hole are described by a choice of a (colored) partition of the product of the number of branes, $N_1N_5$, see \textit{e.g.} \cite{David:2002wn,Balasubramanian:2005qu}. Each partition corresponds to a way to transforming $N_1N_5$ singly-wound strands into a collection of multi-wound strands. The entropy is dominated by microstates that are made of multi-wound strands of length $\leq \sqrt{N_1N_5}$, so it would make sense to think of the our $\sqrt{q_0p^1}$ states as the multi-wound strands of length $\leq \sqrt{N_1N_5}$. However, this description of the microstates is done at the free orbifold point, where our original system with $N_1$ D1-brane charge and $N_5$ D5-brane charge has been mapped through a series of duality transformations to a system with $N_1N_5$ D1 branes and 1 D5 brane \cite{Larsen:1999uk}. To understand what the $\sqrt{N_1N_5}$ microstates are in our original duality frame, one has to follow them back through the duality chain. We hope to report on this in the future.

\section{Conclusions}
\label{sec:conclusions}

In this work we have argued that classical black holes built using the 2-derivative action know about UV scales: the smallest representative in families of black holes that drag the moduli to the boundary of moduli space often follow the species scale or a KK scale. This implies a non-trivial interplay of very different energy regimes, reminiscent of UV/IR mixing. We have carried out this exploration for 5d and 4d BPS black holes, in a systematic analysis of the different infinite distance limits. We have described the corresponding UV scales explored by the minimal classical black holes, showing they reproduce KK scales in decompactification limits or the species scale, even in emergent string limits. We have also provided heuristic arguments for these remarkable facts, using the role of diverging charges in the microstate structure of such black holes, and their interplay with the asymptotic towers of states. We have moreover shown in several illustrative examples that the agreement persists, albeit at a semiquantitative level, in the interior of moduli space. Finally, we have explored the use of the species length bound on classical black hole sizes as a criterion to rule out inconsistent EFTs, using a 9d bottom-up model arising from an  {\em ad hoc} truncation of maximal supergravity.

Some interesting open directions are:

\begin{itemize}

\item The concept of emergence seems to play a key role in providing a rationale for several swampland conjectures. In our particular setup, the emergence of tree level 2-derivative terms seems the key UV/IR mixing ingredient underlying the capability of classical black holes to probe UV scales, oftentimes the species scale. It would be interesting to make further progress in connecting our arguments with the emergence of tree level 2-derivative terms in other contexts.

\item We have checked our ideas mostly in the context of charged BPS black holes, and have shown that they can recover the species scale or other KK related UV scales, in different dimensions. It would be interesting to develop these ideas further and to apply them to also uncharged black holes along the lines of \cite{Bedroya:2024uva}.

\item 
Our results imply that, for large classes of infinite distance limits, there are two ways to define the species scale using black holes. On one hand, one may build the minimal black holes by starting from zero-size black holes and puffing them up to develop an species-scale sized stretched horizon via higher curvature corrections. In CY compactification examples, these corrections typically involve second Chern classes of the Calabi-Yau. On the other hand, we have seen that in certain infinite distance limits, the species scale may be obtained from classical black holes built from the two-derivative action. This requires only knowledge about the triple intersection numbers of the Calabi-Yau. Agreement between the two implies a non-trivial correlation between different topological quantities for Calabi-Yau threefolds, which to our knowledge has not been noticed in the literature. It would be interesting to unravel this mathematical relation and its other possible physical implications.

\item  Our use of minimal classical black holes has led us to the introduction of a new convex hull picture for the explored UV scales, which we have explicitly computed for the bottom-up 9d toy model, given its simplicity. It would be  interesting to study the black hole convex hull picture in a systematic way in more general setups such as 4d ${\cal N}=2$ theories, and to clarify its interplay with the convex hull formulation of other swampland conjectures.

\item 
The 4d and 5d black holes we have explored display an attractor mechanism for vector multiplets, allowing them to explore infinite distance limits for the vector moduli space. Hypermultiplets are decoupled from the attractor flow and remain as independently fixed moduli in the process. There are several possible avenues to include hypermultiplets in the discussion, and explore the corresponding infinite distance limits. For instance, one may use the same black holes as in the present work, and scale the free hypermultiplet vevs together with the black hole charges, so as to reach combined asymptotic limits. Alternatively, one may include additional charges associated to geometric fluxes in the correspondence in \cite{Cribiori:2023swd} to build black holes with less supersymmetry, in which both vector and hyper moduli are fixed at the horizon.

\item We saw that in 4d, although there are black hole families that follow the species scale for the STU model, once additional ``spectator'' moduli are included, the black holes that probe the same limit are necessarily larger than the species scale. It would be interesting to understand more systematically how subleading terms in the prepotential affect the black hole solutions and whether or not they are sensitive to UV scales.

\item We have shown that in a 9d bottom-up toy model it is possible to find black hole families with sizes parametrically smaller than the species scale of their naive UV completion, and that this violation of our proposal signals missing ingredients in this completion, in particular axion partners for the scalar moduli. It would be interesting to study the general applicability of our proposed bound on classical black hole sizes as a swampland criterion to rule out EFTs in general setups.

We hope to come back to these and other interesting questions in the future.

\end{itemize}

\hypersetup{
    pdftitle={Title},
    pdfauthor={blabla},
    pdfsubject={blabla}
}

\section*{Acknowledgements}
We are very grateful to Roberta Angius, Alberto Castellano, \'Alvaro Herr\'aez, Jes\'us Huertas, Tom\'as Ort\'in and Matteo Zatti for useful discussions and for collaborations on related topics. We also thank Roberto Emparan for many interesting discussions and insightful comments. We thank the Erwin Schr\"odinger International Institute for Mathmatics and Physics for their hospitality during the programme “The Landscape vs. the Swampland”. The work of YL is supported by the University of Padua under the 2023 STARS Grants@Unipd programme (GENSYMSTR – Generalized Symmetries from Strings and Branes) and in part by the Italian MUR Departments of Excellence grant 2023-2027 "Quantum Frontiers”. The work of DL is supported by Origins Excellence Cluster and by the German Research Foundation through a German-Israeli Project Cooperation (DIP) grant ``Holography and the Swampland''. AU acknowledges the support of the grants CEX2020-001007-S and PID2021-123017NB-I00, funded by MCIN/AEI/10.13039/501100011033 and by ERDF A way of making Europe.

\newpage

\appendix

\bibliographystyle{JHEP}
\bibliography{draft}

\end{document}